\documentclass[a4paper,12pt]{article}

\usepackage{jheppub} 

\usepackage{fancyhdr}
\usepackage{graphicx}
\usepackage{color}

\usepackage{slashed}

\newcommand{\beq}{\begin{eqnarray}}
\newcommand{\eeq}{\end{eqnarray}}
\newcommand{\nn}{\nonumber}
\def\ltap{\ \raise.3ex\hbox{$<$\kern-.75em\lower1ex\hbox{$\sim$}}\ }
\def\gtap{\ \raise.3ex\hbox{$>$\kern-.75em\lower1ex\hbox{$\sim$}}\ }

\def\eps{\epsilon}

\def\be{\begin{equation}}
\def\ee{\end{equation}}
\def\bea{\begin{eqnarray}}
\def\eea{\end{eqnarray}}

\newcommand{\hc}{\rm h.c.}
\definecolor{red1}{cmyk}{0,1,1,0.3}

\title{\boldmath  The Dark Side of Electroweak Naturalness Beyond the MSSM }

\author[a]{Genevi\`eve B\'elanger,}
\author[a]{C\'edric Delaunay,}
\author[a,b]{Andreas Goudelis}
\affiliation[a]{LAPTh, Universit\'e de Savoie, CNRS, 9 Chemin de Bellevue, B.P.\
110,
F-74941 Annecy-le-Vieux, France
}
\affiliation[b]{Institute of High Energy Physics, Austrian Academy of Sciences, \\
Nikolsdorfergasse 18, 1050 Vienna, Austria
}
\emailAdd{genevieve.belanger@lapth.cnrs.fr}
\emailAdd{cedric.delaunay@lapth.cnrs.fr}
\emailAdd{andreas.goudelis@oeaw.ac.at}

\abstract{
Weak scale supersymmetry (SUSY) remains a prime explanation for the radiative stability of the Higgs field. 
A natural account of the Higgs boson mass, however, strongly favors extensions of the Minimal Supersymmetric Standard Model (MSSM). A plausible option is to introduce a new supersymmetric sector coupled to the MSSM Higgs fields, whose associated states resolve the little hierarchy problem between the third generation squark masses and the weak scale. SUSY also accomodates a weakly interacting cold dark matter (DM) candidate in the form of a stable neutralino. In minimal realizations, the thus-far null results of direct DM searches, along with the DM relic abundance constraint, introduce a level of fine-tuning as severe as the one due to the SUSY little hierarchy problem. We analyse the generic implications of new SUSY sectors parametrically heavier than the minimal SUSY spectrum, devised to increase the Higgs boson mass, on this “little neutralino DM problem”. We focus on the SUSY operator of smallest scaling dimension in an effective field theory description, which modifies the Higgs and DM sectors in a correlated manner.
Within this framework, we show that recent null results from the LUX experiment imply a tree-level fine-tuning for gaugino DM which is parametrically at least a few times larger than that of the MSSM. Higgsino DM whose relic abundance is generated through a thermal freeze-out mechanism remains also severely fine-tuned, unless the DM lies below the weak boson pair-production threshold. As in the MSSM, well-tempered gaugino-Higgsino DM is strongly disfavored by present direct detection results.
}

\begin{document}
\maketitle
\flushbottom

\section{Introduction} \label{sec:intro}

Persuasive gravitational evidence from the galactic scale and above suggests that our Universe is filled with an unknown form of non-baryonic, dark matter (DM) (see {\it e.g.} Refs.\cite{JKGdmrev,CMdmrev,BHSdmrev} for a review). Despite these observations, very little is known about the nature of the DM as well as its non-gravitational properties. A very attractive possibility is that DM is a cosmological relic in the form of non-relativistic, collisionless particles. Even within this paradigm the  range of possible DM mass scales is very broad and the DM interactions are not specified. The thermal freeze-out mechanism for generating the DM relic density is of particular interest as it
suggests that DM particles couple to Standard Model (SM) fields, thus opening the possibility to probe the dark sector through known interactions other than gravity. 
It further offers the possibility to connect DM to the weak scale since an $\mathcal{O}(100\,$GeV) DM particle whose couplings to SM fields are comparable in strengh to the SM weak ones, freezes out with a relic density of the right order of magnitude. This is the so-called weakly interacting massive particle (WIMP) miracle.
The same interactions would make relic DM particles in our immediate neighborhood directly visible through their scattering on nuclei~\cite{WittenDD}, as well as allow for DM production at high energy colliders~\cite{Baer:2009bu}, provided its mass is not too large. 
Despite the remarkable efforts of direct detection experiments~\cite{XENON10,XENON100,CDMS2,LUX,CRESST2} and the completion of the $8\,$TeV LHC run, DM particles with properties consistent with the WIMP miracle remain elusive.

On completely different scales, the recent discovery~\cite{ATLASdisco,CMSdisco} of a $\simeq125\,$GeV Higgs boson at the LHC also calls for the existence of new particles beyond the SM. A light SM Higgs is subject to a severe hierarchy problem which requires either an unnaturally large fine-tuning of seemingly unrelated SM parameters or a new structure to emerge not far below the TeV scale in order to screen  the weak scale from large radiative corrections at very short distance. Although naturalness of the Higgs mass does not {\it a priori} predict the existence of a particle stable on cosmological scales, the coincidence of the plausible mass scales for DM and naturalness-motivated new physics remains intriguing. TeV-scale supersymmetry (SUSY) is a well motivated solution of the hierarchy problem which can easily accommodate a DM candidate.  The lightest SUSY particle (LSP), if colorless and electrically neutral, gathers the required basic properties to act as DM, provided its decay back to SM states is forbidden by a sufficiently well-preserved $R$-parity symmetry~\cite{JKGdmrev}.
The canonical candidate with the above properties is the lightest neutralino, {\it i.e.} the lightest SUSY partner of the neutral SM electroweak (EW) states. Neutralino DM scenarios are particularly interesting as their phenomenology is directly tied to the Higgs sector. This connection constitutes one of the rare occasions where DM affects EW naturalness.\\

Within the above framework, the minimal SUSY extension of the SM (MSSM) with exact $R$-parity is the most economical way to address both EW scale naturalness and DM. However, it is not possible in this model to accommodate a Higgs boson mass as large as $\simeq125\,$GeV without a sizable source of SUSY breaking in the top quark/squark sector, which, rather ironically,  reintroduces
 a percent-level sensitivity of the weak scale to arbitrarily short distance dynamics~\cite{HallPinnerRuderman}. This defines the SUSY little hierarchy problem. Its resolution motivated various extensions of the MSSM, most of which invoke the existence of new light degrees of freedom around the MSSM ones. 
Those include in particular the addition of a gauge singlet superfield (NMSSM)~\cite{ManiatisNMSSM,EllwangerNMSSM} or extra (spontaneously broken) gauge groups ~\cite{BatraDterms,MaloneyDterms,AuzziGiveon,WagnerSU2,BGMcGDterms,KielScherkSchwarz}, both offering the possibility of a reduced fine tuning as compared to the MSSM ~\cite{Ellwanger:2011mu,Ross:2012nr,Binjonaid:2014oga,Kaminska:2014wia,Athron:2013ipa,McGarrie:2014xxa}.
Another attractive approach consists in introducing a new SUSY sector slightly decoupled from the MSSM degrees of freedom. The separation of scales then allows for an effective treatment of the new sector beyond the MSSM. This SUSY effective approach is referred to in the literature as the BMSSM~\cite{DST}. The authors of Refs.~\cite{Strumia,CasasEspinosaHidalgo,DST} pointed out that the leading higher-dimensional operator in the Higgs sector could bring the Higgs boson mass to its observed value, provided the BMSSM scale is within a few TeV.\footnote{Higher-dimensional operators could even dominate the Higgs boson mass prediction within the range of validity of the  effective field theory. This feature results from the fact that the tree-level Higgs quartic interactions are doubly suppressed in the MSSM by small EW gauge couplings and the presence of $D$-flat directions.}

Although the large amount of SUSY breaking in the top sector typically constitutes the dominant source of MSSM fine-tuning, another important source arises from the SUSY-preserving $\mu$ parameter controlling the Higgsino masses. If DM is to be identified with the lightest neutralino, direct detection searches and/or relic density constraints yield a unique probe of  the fine-tuning associated with the Higgsino decoupling, potentially more efficient than direct electroweakino searches at the LHC. Direct detection typically forces the LSP to project almost entirely on either gaugino or Higgsino states, thus suppressing the dominant Higgs exchange amplitude. Having a gaugino LSP requires decoupling the $\mu$ parameter which, hence, induces unacceptably large fine-tuning. In contrast,  Higgsino LSP satisfies direct detection constraints at low fine-tuning provided $\mu$ remains small. It is, however, not possible in this case to recover the observed DM relic density due to very efficient LSP annihilation and co-annihilation processes, unless the DM mass is sufficiently large. This implies a large $\mu\sim \mathcal{O}($TeV) and again too large fine-tuning. The authors of Ref.~\cite{Perelstein1} showed that the Xenon100 results already raise the tree-level fine-tuning in the MSSM up to the percent-level, which is of the same order as the fine-tuning originating from heavy third generation squarks (and the gluino). 
In particular,  this implies that neutralino DM searches in direct detection experiments constitute a complementary probe of weak scale naturalness, potentially more efficient  than top squark and gluino searches at the LHC. Moreover, the overall fine-tuning level may remain significant, through a dominant Higgsino source, in MSSM extensions which otherwise solve the little hierarchy problem. This is the case for instance in the NMSSM, unless the $\lambda$-SUSY limit is assumed~\cite{Perelstein2}.\\

The main goal of the present paper is to study the implications of DM phenomenology on EW naturalness in the BMSSM. The effective field theory (EFT) nature of this framework allows a generic analysis of such a DM/naturalness connection in MSSM extensions where the little hierarchy problem is solved through an extra heavy SUSY sector. In particular, we find interesting that, due to its SUSY-preserving nature, the leading BMSSM operator in the Higgs sector modifies the Higgsino properties in a way which completely correlates with the Higgs mass, provided SUSY-breaking contributions in the top/stop sector are small as required by naturalness. There are existing studies in the literature on the BMSSM neutralino dark matter relic density~\cite{BMSSMedsjo,BMSSMnir} and direct detection prospects~\cite{BMSSMdetection}.  However, to the best of our knowledge, none of them attempted to connect neutralino DM phenomenology to the question of weak scale naturalness in this framework.
We first update the MSSM results of Ref.~\cite{Perelstein1} by taking into account the recent null results of the LUX experiment~\cite{LUX}, as well as assuming more up-to-date estimates for the hadronic parameters entering the spin-independent (SI) neutralino-nucleon scattering cross section. For  gaugino LSP, we show that direct detection constraints always imply a significantly larger Higgsino fine-tuning in the BMSSM relative to the MSSM. Using current data from the LUX experiment, the fine-tuning is worsened by a factor of up to $4$ for LSP masses around $30-50\,$GeV. 
For  Higgsino LSP, on the other hand, the level of fine-tuning remains comparable to that of the MSSM whenever the (co-)annihilation channels into weak bosons are open. We find, however, that the BMSSM operator is critical in obtaining the observed DM relic density for Higgsino LSP below the weak boson pair-production threshold, while keeping the charginos above the kinematic LEP bound. This is the only region of parameter space with a moderately low fine-tuning which is consistent with collider, direct DM detection and DM relic abundance constraints. A smoking gun signature of this scenario is a light Higgsino-like chargino state just above the kinematic LEP2 bound, $m_{\tilde C}\gtrsim 103\,$GeV.\\

The remainder of the paper is organized as follows. In Sec.~\ref{sec:BMSSMreview} we review the EFT description of the leading BMSSM operators, and their effects on the SM Higgs mass. Their impact on the tree-level source of EW fine-tuning is analyzed in Sec.~\ref{sec:FT}, while in Sec.~\ref{sec:neumix} we review the associated modifications in the neutralino and chargino sectors. In Sec.~\ref{sec:darkmatter} we analyse the implications of  direct DM searches and/or the relic density on EW fine-tuning in the BMSSM in comparison with the renormalizable MSSM, whenever possible. We present our conclusions in Section~\ref{sec:outro}.

\section{Effective description of new physics beyond the MSSM} 
\label{sec:BMSSMreview}
We assume the MSSM is extended by a new supersymmetric sector whose characteristic mass scale $M$ is parametrically larger than that of the MSSM states, collectively denoted by $m_{\rm soft}$. The dynamics of such heavy supersymmetric sectors is then well described by an effective superpotential $\mathcal{W}_{\rm eff}$ whose least irrelevant operator involving only Higgs superfields is~\cite{Strumia,DST}
\be\label{ops1}
\mathcal{W}_{\rm eff} = \mu H_u\cdot H_d + \frac{\lambda_1}{M}(H_u\cdot H_d)^2+\cdots\,,
\ee
where the ellipses denote MSSM Yukawa interactions and $\mathcal{O}(1/M^2)$ and higher operators. 
$H_{u,d}$ are the chiral superfields of the Higgs doublets and  $H_u\cdot H_d=H_u^T(i\sigma_2)H_d$ denotes their antisymmetric product.
There are operators at $\mathcal{O}(1/M)$ which couple the Higgs with other chiral superfields in the Kahler potential, {\it e.g.} $\int d^4\theta H_d^\dagger Q u^c+\hc$. However, those are irrelevant to our analysis as they contribute neither to the Higgs spectrum nor to the Higgs-to-neutralino couplings. There are also additional operators which violate baryon and/or lepton number~\cite{DRT,GKM}. It is reasonable to assume that the underlying baryon and lepton number breaking dynamics arises at a much higher scale than $M\sim\mathcal{O}($few TeV). For these reasons we only consider the $\mathcal{O}(1/M)$ operator of Eq.\eqref{ops1} (see Refs.~\cite{BMSSM6,Boudjema:2011aa,Boudjema:2012cq,Boudjema:2012in} for BMSSM analyses including dimension six operators in the Higgs sector).
 
Once SUSY breaking is mediated to the effective theory, the following soft Lagrangian is induced
\beq
\mathcal{L}^{\rm soft} = \mathcal{L}^{\rm soft}_{\rm MSSM} + \int d^2\theta\  \frac{\lambda_2}{M}X (H_u\cdot H_d)^2+\hc\,,\label{ops2}
\eeq
where $X=m_{\rm soft}\theta^2$ is a dimensionless $F$-term spurion parameterizing SUSY breaking effects~\footnote{We assume here that $D$-term breaking effects are subdominant~\cite{ZoharDbreaking}.}. The MSSM soft terms are
\beq
-\mathcal{L}^{\rm soft}_{\rm MSSM} &=& m_{H_u}^2 |h_u|^2+m_{H_d}^2|h_d|^2+(b\, h_u\cdot h_d+\hc)\nn\\
&&+ \frac{M_1}{2}\tilde B\tilde B+\frac{M_2}{2}\tilde W^a \tilde W^a + \cdots\,,
\eeq
where $h_{u,d}$ are the scalar components of $H_{u,d}$, $\tilde W^a$ and $\tilde B$ are the SU(2)$_L\times$U(1)$_Y$ gaugino fields and $\cdots$ denotes the gluino mass, scalar fermion masses and trilinear interaction terms which do not play an important role here. 

The effective operators in Eqs.~\eqref{ops1} and \eqref{ops2} induce new quartic interactions in the Higgs scalar potential
\beq
2\eps_1 (h_u\cdot h_d)\left(|h_u|^2+|h_d|^2\right)+\eps_2(h_u\cdot h_d)^2+\hc\,,
\eeq
as well as extra Higgs-Higgsino interactions
\beq\label{Lhino}
-\frac{\eps_1}{\mu^*}\Big[2(h_u\cdot h_d)(\tilde h_u\cdot \tilde h_d)+2(\tilde h_u\cdot h_d)(h_u\cdot \tilde h_d)
+(h_u\cdot \tilde h_d)^2+(\tilde h_u\cdot h_d)^2\Big]+\hc\,,
\eeq
where $\tilde h_{u,d}$ are the Higgsino doublets, SUSY partners of $h_{u,d}$ and we defined $\eps_1\equiv \lambda_1\mu^*/M$ and $\eps_2\equiv -\lambda_2m_{\rm soft}/M$. There are four  independent CP phases in Eqs.~\eqref{ops1} and \eqref{ops2} which can be parameterized as
$\arg\left(\mu M_{1,2}/ b\right)$, $\arg\left(\eps_1/b\right)$, and $\arg\left(\eps_2/b^2\right)$~\cite{DimThomas,BMSSMCP}. Some combinations of those phases are typically strongly constrained by electric dipole moment (EDM) searches. Although it is possible to evade EDM constraints for moderate values of the BMSSM phases\footnote{The BMSSM phases could be large enough to drive successful EW baryogenesis in the early Universe~\cite{BMSSMCP}.}, in the following for simplicity we assume CP conservation and set these phases to zero. This assumption is of mild importance as the Higgs spectrum is only corrected by the real part of $\eps_{1,2}$ at leading order~\cite{DST}. We are however left with possible relative signs between the $\mu$-parameter and the gaugino masses $M_{1,2}$. We choose to work in a basis where $\mu>0$ while $M_{1,2}$ could have either sign.

Finally, EW symmetry breaking occurs through the usual interplay between the quadratic and quartic terms in the scalar potential (see Ref.~\cite{PontonBatra} for an alternative scenario). 
We parameterize the resulting vacuum expectation values (VEV) of $h_{u,d}$ as
\beq\label{EWvac}
\langle h_u\rangle=\left(\begin{array}{c} 0 \\ v\sin\beta  \end{array}\right)\,,\quad \langle h_d\rangle=\left(\begin{array}{c} v\cos\beta \\ 0  \end{array}\right)\,,
\eeq
with $v\simeq 174\,$GeV and $0\leq \beta\leq \pi/2$. 

\subsection{Higgs boson mass}\label{sec:HBmass}
Around the vacuum of Eq.~\eqref{EWvac} the mass of the lightest CP-even Higgs boson $h$ is 
\be\label{Hmass}
m_h^2 = m_{h,0}^2+\delta_\eps^2+\delta_{\rm rad}^2,
\ee
where $m_{h,0}^2\leq m_Z^2$ is the tree-level MSSM prediction, $\delta_{\rm rad}^2$ represents radiative corrections dominated by top/stop loops, and~\cite{DST}
\be\label{dmh2eps}
\delta_\eps^2 =2v^2\left(\eps_2-2\eps_1 \sin2\beta -\frac{2\eps_1 x \sin2\beta+\eps_2y\cos^22\beta}{\sqrt{y^2+(x^2-y^2)\sin^22\beta}}\right) \,
\ee
is the leading tree-level correction arising from the effective operators in Eqs.\eqref{ops1} and \eqref{ops2}.
In Eq.\eqref{dmh2eps}, we defined $x=m_A^2+m_Z^2$ and $y=m_A^2-m_Z^2$, where  $m_A$ denotes the CP-odd Higgs mass. The mass of the heavy CP-even Higgs scalar as well as the angle $\alpha$ setting the orientation of mass eigenstates relative to the vacuum are also corrected at $\mathcal{O}(\eps)$. We refer the reader to Appendix~\ref{Hspectrum} for further details.\\

In the MSSM ($\delta_\eps=0$), $m_h^2\simeq (125\,{\rm GeV})^2$ is only obtained at the expense of radiative corrections almost as large as the tree-level contribution $\delta_{\rm rad}\sim m_{h,0}^2$ and for large $\tan\beta$. This implies large SUSY-breaking soft terms for the third generation squarks, $m_{\rm soft}\gtrsim \mathcal{O}(1\,$TeV$)$, which by itself reintroduces a fine-tuning of the EW scale at the percent level or worse~\cite{HallPinnerRuderman,ArvanitakiVilladoro}.
\begin{figure}[!t]
\begin{center}
\begin{tabular}{c}
\includegraphics[width=0.6\textwidth]{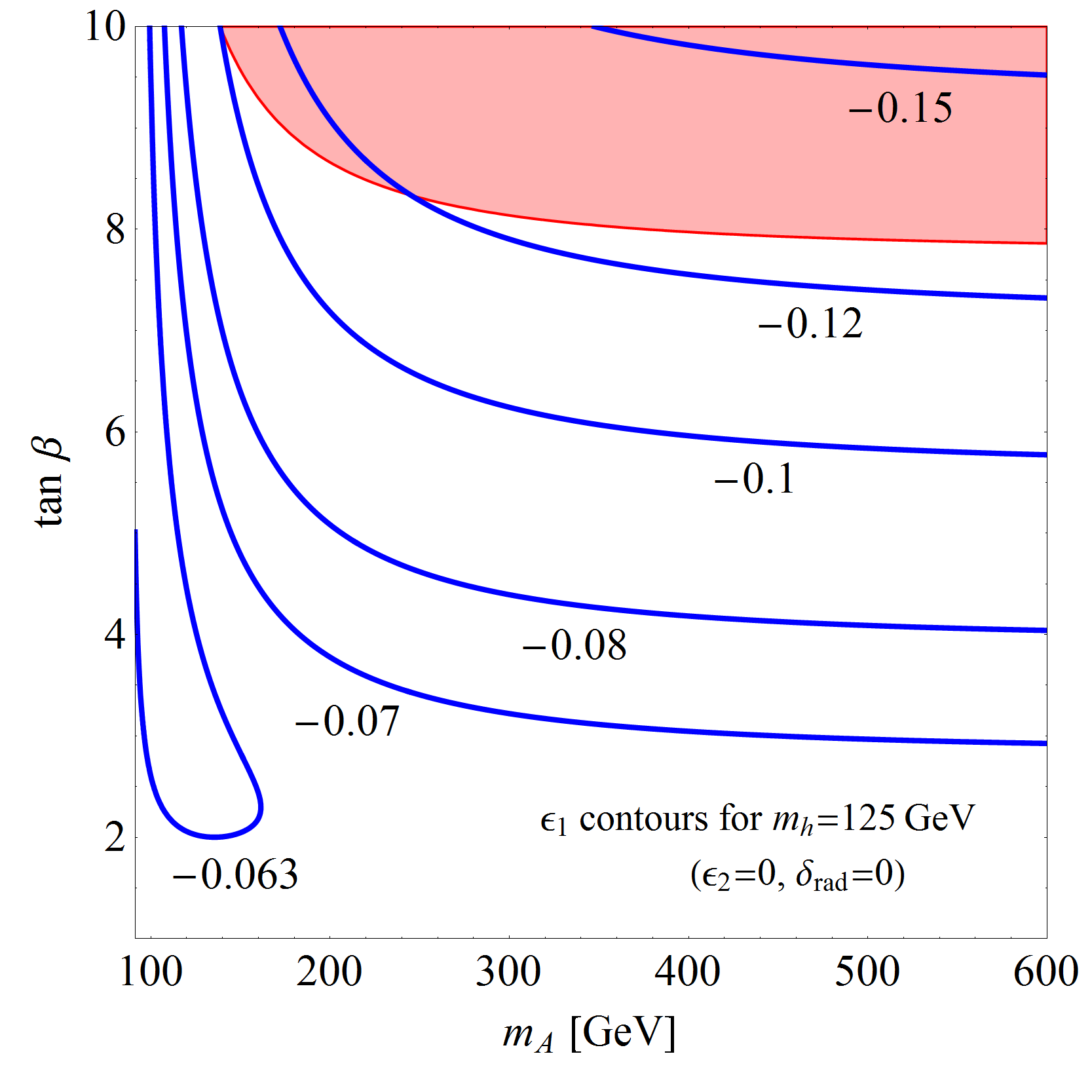}\\
\includegraphics[width=0.6\textwidth]{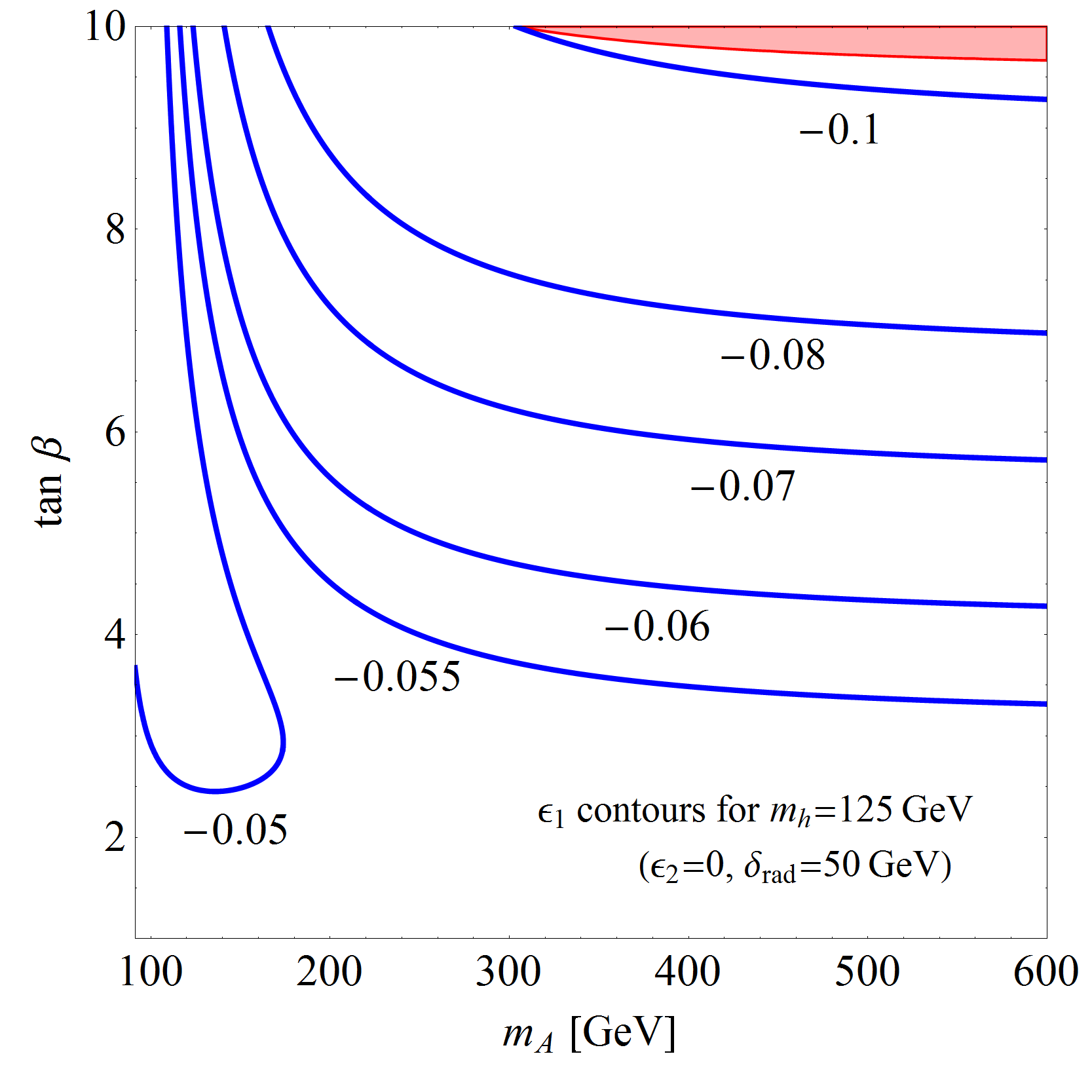}
\end{tabular}
\caption{Contours of $\eps_1$ values required in order to obtain $m_h=125\,$GeV as function of $m_A$ and $\tan\beta$, for $\epsilon_2=0$ and assuming (top) $\delta_{\rm rad}^2=0$ or (bottom) $\delta_{\rm rad}^2=(50\,$GeV$)^2$. The region where the Higgs mass correction is no longer dominated by the leading order effect from the dimension five operator in Eq.~\eqref{ops1} ($|\eps_1|\tan\beta\geq1$) is shown in red.}
\label{epscontours}
\end{center}
\end{figure}
For a relatively low BMSSM scale $M\simeq \mathcal{O}($few$\,$TeV), this tension can be significantly relaxed. Furthermore, for $-\eps_1\simeq \mathcal{O}(0.1)$ it is even possible to accommodate a 125$\,$GeV Higgs already {\it at tree-level}~\cite{DST}, which corresponds to a BMSSM scale of $M\simeq 1\,$TeV$\times \lambda_1(\mu/100\,$GeV$)$\footnote{The range of validity of the EFT could be pushed to $M\sim\mathcal{O}(10\,$TeV) if the new sector is strongly coupled at the cut-off with $\lambda_1\sim 4\pi$.}. 
Direct searches at the LHC limit lightest stop masses to values which strongly depend on the LSP mass. Current limits are as high as $m_{\tilde t}\gtrsim 670\,$GeV for a massless LSP, and weaken for a heavier LSP~\cite{Aad:2014kra,Chatrchyan:2013xna}.\footnote{Top squarks close to kinematic thresholds yield too soft decay products and cannot be excluded by direct LHC searches. These stealthy regions~\cite{stealthstop} can nevertheless be probed by either precise cross section~\cite{MitovWeiler} or spin-spin correlation~\cite{spincorrstop} measurements in top pair production.} For a $\simeq150\,$GeV LSP, top squarks as light as $\simeq 300\,$GeV are allowed.
The value of $\eps_1$ needed to bring the Higgs mass prediction in the BMSSM at the observed value varies as a function of $\tan\beta$ and $m_A$. Figure~\ref{epscontours} illustrates this dependence, as dictated by Eq.~\eqref{dmh2eps}, for $\delta_{\rm rad}^2=0$ and $\delta_{\rm rad}^2=(50\,$GeV$)^2$, corresponding to unmixed degenerate top squarks of $\mathcal{O}(300\,$GeV) mass, respectively. Because of its SUSY-breaking origin the $\eps_2$ effect is parametrically subdominant relative to that of $\eps_1$, as easily appreciable in the decoupling limit $m_A\gg m_Z$ where
\be
\delta_\epsilon^2\simeq -8 v^2\left(\eps_1\sin2\beta-\frac{\sin^22\beta}{4}\eps_2\right)+\mathcal{O}\left(\frac{m_Z^2}{m_A^2}\right)\,.
\ee 
For instance, taking $m_A\simeq 300\,$GeV and $\tan\beta\simeq3$, the $\eps_2$ value  required to obtain the correct Higgs mass at tree-level (assuming $\eps_1=0$) is a factor $\simeq 4\tan\beta\sim\mathcal{O}(10)$ larger than that of $\eps_1$ (assuming $\eps_2=0$).  Note also that both $\mathcal{O}(\eps)$ effects are suppressed at large $\tan\beta$.\footnote{This is in contrast with, for instance, gauge extensions of the MSSM which enhance the Higgs mass through non-decoupled $D$-terms~\cite{BatraDterms,MaloneyDterms,AuzziGiveon,WagnerSU2,BGMcGDterms}, a contribution of SUSY-breaking origin which  increases with $\tan\beta$.}
 Higher orders typically do not suffer from such a suppression~\cite{BMSSM6}. Therefore, for sufficiently large $\tan\beta$,  the Higgs mass correction is no longer dominated by $\mathcal{O}(\eps)$ effects.  This signals a lack of predictivity of the EFT with regards to the light CP-even Higgs mass. We therefore choose to restrict our analysis to $\tan\beta$ values low enough so that the EFT prediction in Eq.~\eqref{Hmass} at $\mathcal{O}(\eps_1)$ for $m_h$ is reliable, which is the case for $|\eps_1|/\tan\beta \gtrsim \eps_1^2$ or, equivalently, 
\be\label{tbmax}
\tan\beta \lesssim |\eps_1|^{-1}\sim \mathcal{O}(10)\,. 
\ee
This is in contrast with the renormalizable MSSM where much larger $\tan\beta$ values are allowed. 
Since the $\eps_2$ contribution remains negligibly small whenever the EFT description is valid and does not correlate with DM observables through the Higgsino sector, we choose to ignore it and set $\eps_2=0$. \\

The effective operator in Eq.~\eqref{ops1} also induces a second (remote) vacuum at $\langle h_u^0\rangle\simeq\langle h_d^0\rangle\sim \sqrt{\mu M}\ll M$, in the presence of which the EW vacuum of Eq.~\eqref{EWvac} may be unacceptably short-lived~\cite{BDH}. Stability of the EW vacuum along the dangerous $D$-flat direction is guaranteed under the condition (assuming $\eps_2=0$)~\cite{BDH}
\be\label{stab}
 \mu\lesssim m_A\sqrt{\frac{1+\sin2\beta}{2}}\left[1+\frac{8v^2}{m_A^2}\left(\frac{1+2\sin2\beta}{1+\sin2\beta}-\frac{3}{2}\right)\right]^{1/2}\,,
\ee
which we shall assume true in this paper. Strictly speaking, a mild violation of this condition is still allowed as a meta-stable EW vacuum remains phenomenologically viable provided its lifetime exceeds the age of the Universe. A careful analysis of the tunneling rate reveals that the condition in Eq.~\eqref{stab}, besides being more practical, is rather accurate and slightly conservative~\cite{BDH}.
Away from this $D$-flat direction, the MSSM $D$-terms stabilize the EW vacuum provided $\eps_1^2\lesssim m_Z^2/4v^2$~\cite{BDH}, {\it i.e.} $|\eps_1|\lesssim0.25$, which, according to Fig.~\ref{epscontours}, is always fulfilled whenever Eq.~\eqref{tbmax} holds.

\section{BMSSM Electroweak Fine-tuning} \label{sec:FT}
The $Z$ boson mass and $\tan\beta$ are set by the minimization conditions of the scalar potential assuming the vacuum in Eq.~\eqref{EWvac}. To leading order in $\eps_{1,2}$, we find the tree-level relations
\be\label{mZrel}
m_Z^2=
\frac{|m_{H_d}^2-m_{H_u}^2|}{\sqrt{1-\sin^22\beta}}-m_{H_u}^2-m_{H_d}^2-2\mu^2+4\eps_1 v^2\sin2\beta\,,
\ee
and
\be
\sin2\beta=\frac{2b}{m^2}+\frac{4v^2}{m^2}\left[\eps_1\left(1+4\frac{b^2}{m^4}\right)-\eps_2\frac{b}{m^2}\right]\,,
\ee
where $m^2\equiv m_{H_u}^2+m_{H_d}^2+2\mu^2$. The stability of the EW scale well below the cutoff scale is threatened whenever some mass parameters in Eq.~\eqref{mZrel} take values much larger than $m_Z$ unless an unnatural cancellation among these parameters occurs. 
We quantify the amount of fine-tuning associated with a
 model's parameter $p$ through a Barbieri--Giudice measure\footnote{Fine-tuning measures are  subjective to some extent, and the resulting estimates are not particularly sharp quantitatively. However, the \textit{difference} of fine-tuning between two sets of model parameters is a more physically robust quantity.}~\cite{BG}
\beq\label{FTindiv}
\Delta_p \equiv \left|\frac{\partial \log m_Z^2}{\partial\log p}\right|\,.
\eeq
Under the assumption that all $\Delta_p$'s are independent, a global measure of fine-tuning is obtained by summing them in quadrature
\be\label{FTglobal}
\Delta \equiv \sqrt{\Delta_{0}^2+\Delta_{\rm rad}^2}\,,\quad \Delta_0\equiv \sqrt{\sum_p \Delta_p^2}
\ee 
where the sum runs over $p=\mu,b,m_{H_u}^2,m_{H_d}^2,\eps_1,\eps_2$. $\Delta>1$ means an overall fine-tuning of $1/\Delta$. $\Delta_{\rm rad}$ parameterizes the fine-tuning associated with the set of MSSM parameters which only contribute to the relation Eq.~\eqref{mZrel} at loop level, of which the stop quark masses and mixing parameter (and to lesser extend the gluino mass) are the most relevant.\\ 

Within the MSSM, $m_h\simeq 125\,$GeV requires large stop masses and/or mixing which enter  Eq.~\eqref{mZrel} quadratically through one-loop renormalization of the Higgs soft masses. As argued in section \ref{sec:HBmass}, large SUSY-breaking effects are no longer necessary in the top/stop sector in the presence of the higher dimensional operator in Eq.~\eqref{ops1}.
The overall fine-tuning is then dominated by the relative sensitivity of $m_Z^2$ to the tree-level parameters listed above. This tree-level source of fine-tuning typically correlates with DM observables, mostly through the $\mu$-paremeter~\cite{Perelstein1,Perelstein2}. The null results of DM direct detection searches and the thermal relic density already strongly constrain the composition of the lightest neutralino, which in turn implies a non-negligible source of fine-tuning $\Delta_0$.  

The effective operators in Eqs.~\eqref{ops1} and \eqref{ops2} modify the Higgs scalar spectrum and the vacuum, hence $\Delta_0$, in a correlated way. The complete analytical expressions, corrected at $\mathcal{O}(\eps)$, for the $\Delta_p$'s are rather lengthy and can be found in Appendix~\ref{app:DeltaEW}. 

We show in Fig.~\ref{figFT} the relative variation of  $\Delta_0$ between the BMSSM and the MSSM as function of the MSSM $\Delta_0$ for several values of $\tan\beta$ and $m_A$. As clearly apparent the tree-level fine-tuning can be improved, most notably for moderately low $\tan\beta\lesssim4$ and light $m_A\lesssim300\,$GeV. The improvement can reach up to $\sim\mathcal{O}(40\%)$ when $\Delta_0\simeq 20$, which corresponds to $\mu\simeq 100\,$GeV. For larger $\Delta_0$ values, Fig.~\ref{figFT} further illustrates a significant limitation in the fine-tuning improvement in the BMSSM due to the vacuum stability constraint. Equation~\eqref{stab} indeed requires, relative to the MSSM, larger values of $m_A$ for a fixed $\mu$-parameter. The implications of the BMSSM stability constraint are easily understood by expanding the tree-level fine-tuning in the large $\tan\beta$ limit (yet still satisfying Eq.~\eqref{tbmax}).
The dominant fine-tuning sources to leading order in $\eta\equiv \tan^{-1}\beta$  are

\be\label{FTleading}
\Delta_\mu\simeq \frac{4\mu^2}{m_Z^2}\left(1+8\eta\frac{\eps_1v^2}{m_A^2}+\mathcal{O}(\eta^2)\right)\,,
\ee
\be\label{FTleading2}
\Delta_{m_{H_u}^2}\simeq \left(1+\frac{2\mu^2}{m_Z^2}\right)\left[1+4\eta\frac{\eps_1v^2}{ m_A^2}\left(1- \frac{2m_A^2}{m_Z^2+2\mu^2}\right)+\mathcal{O}(\eta^2)\right]\,,
\ee
to leading $\mathcal{O}(\eps_1)$, while
\be
\Delta_b\simeq 2\Delta_{m_{H_d}^2}\simeq \frac{2\eta^2m_A^2}{m_Z^2}\,,
\ee
could also be relevant whenever $m_A\gtrsim\mu\tan\beta$. We first observe from the above expressions that the higher-dimensional operator of Eq.~\eqref{ops1} typically helps in reducing the tree-level fine-tuning whenever effective in bringing $m_h$ up to the observed value at the classical level, {\it i.e.} for $\eps_1<0$. Equations~\eqref{FTleading} and~\eqref{FTleading2} also show that the fine-tuning improvement from the presence of the BMSSM operator is reduced for larger $\tan\beta$. Furthermore, at fixed $\mu$, the large $m_A$ required by vacuum stability tends to suppress the $\epsilon_1$ corrections to the leading fine-tuning contribution $\Delta_\mu$ and $\Delta_{m_{H_u}^2}$, while increasing the sub-leading ones, in particular $\Delta_b$ and $ \Delta_{m_{H_d}^2}$, relative to the MSSM. 
\begin{figure}[t!]
\centering
\begin{tabular}{cc}
\includegraphics[width=0.7\textwidth]{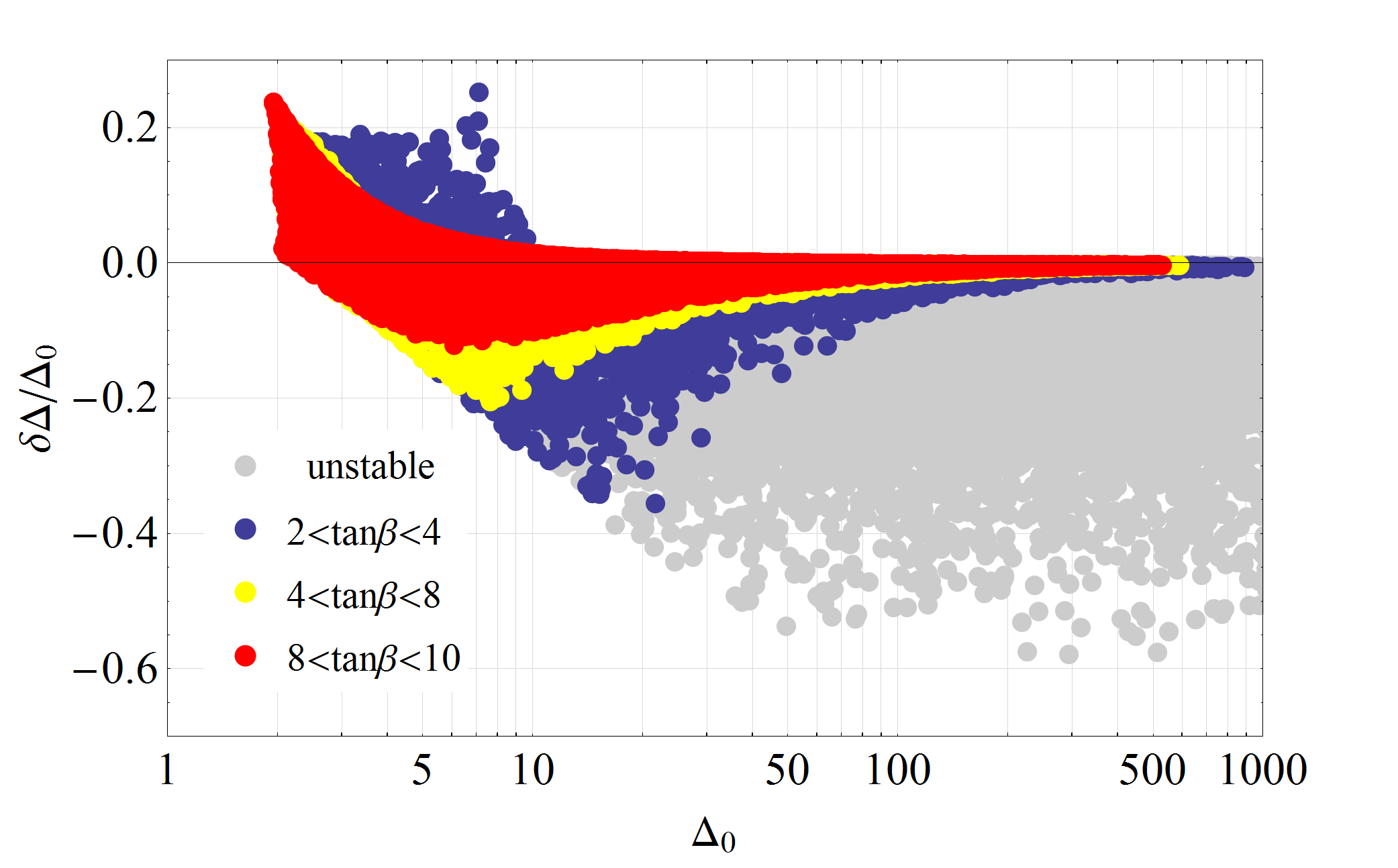}\\
\includegraphics[width=0.7\textwidth]{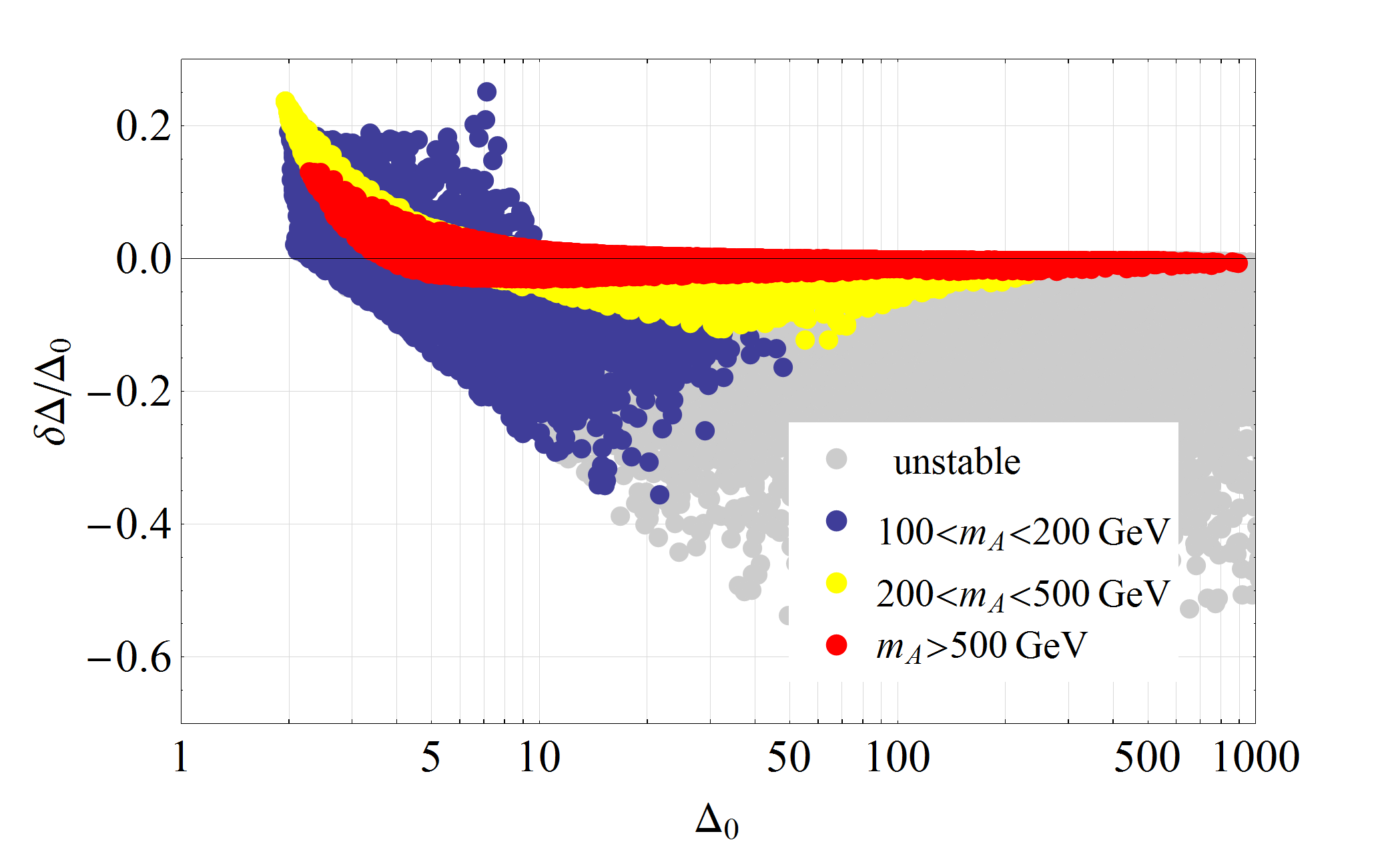}
\end{tabular}
\caption{Relative tree-level fine-tuning variation between the BMSSM and the MSSM with same MSSM parameter values, for various $\tan\beta$ (upper panel) and $m_A$ (lower panel) values. For the BMSSM, the effective operator $\eps_1$ is set in order to obtain $m_h=125\,$GeV  at tree-level with $\epsilon_2=0$. $\delta\Delta<0$ corresponds to an improved fine-tuning relative to the MSSM. Grey points are strongly disfavored as they violate the stability condition of Eq.~\eqref{stab}.}
\label{figFT}
\end{figure}
We finally stress that in the context of neutralino DM, the tension between fine-tuning and direct DM searches is most pronounced in the limit of large $\mu$-parameter where  the LSP is a nearly pure gaugino. However, in this case, as reviewed in the next section, large $\tan\beta$ values suppress the leading amplitude for LSP-nucleon (SI) scattering and thus partially relax the tension with EW naturalness. We therefore do not expect any significant fine-tuning improvement {\it  at fixed} $\tan\beta$ in the BMSSM.

\section{Detection and relic density of neutralino dark matter} \label{sec:neumix}
The effective operator in Eq.~\eqref{ops1} further modifies the neutralino and chargino properties, through the Lagrangian in Eq.~\eqref{Lhino}.
In a natural theory where the stop quarks are light and unmixed,
 these modifications are tightly correlated with the Higgs boson mass through $\eps_1$. 
The lightest neutralino, henceforth denoted $\chi$, is a general admixture of the four current states $\psi^0=(\tilde B, \tilde W^3, \tilde h_d^0, \tilde h_u^0)^T$ and reads
\be
\chi = \mathcal{N}_{\chi k} \psi_k^0\,,\quad \mathcal{N}_{jk}= \eps^{i\phi_j}\mathcal{Z}_{jk} 
\ee
where $\mathcal{Z}$ is the orthogonal matrix diagonalizing the neutralino mass matrix $\mathcal{M}_{\chi_0}$, {\it i.e.} $\mathcal{Z}\mathcal{M}_{\chi_0}\mathcal{Z}^T={\rm diag}(m_\chi\,,\dots)$, and $\phi_\chi=0\ (\pi/2)$ for $m_\chi>0$ ($<0$). 
We evaluate numerically the $\mathcal{O}(\eps_1)$ effect on the lightest neutralino composition. Nevertheless, direct DM searches already strongly disfavor neutralino LSP's which are strong admixtures of gaugino and Higgsino states~\cite{Perelstein1}. Indeed, as shown in Fig.~\ref{DD_compo},  low fine-tuning scenarios ($\mu\sim\mathcal{O}(100\,$GeV)) with significant $\tilde B/\tilde h$ or $\tilde W/\tilde h$ mixing are in tension with direct DM searches at the LUX experiment~\cite{LUX} by one order of magnitude in the neutralino-nucleon scattering cross-section. 
In order to gain insight into the consequences of the BMSSM modifications we derive approximate analytical expressions for the lightest neutralino mass and composition in the cases where $\chi$ is almost a 
\begin{enumerate}
\item  pure bino state, with $M_1\lesssim M_2 \ll \mu$,
\end{enumerate}
or a 
\begin{enumerate}
\setcounter{enumi}{1}
\item pure Higgsino state, with $\mu\lesssim M_{1,2}$. 
\end{enumerate}
Although well-tempered scenarios with a strongly mixed $\tilde B/\tilde W$ LSP are motivated by the relic abundance~\cite{welltempered}, we focus for simplicity on gaugino LSP without wino projection. Since $g>g'$, the latter would lead to a larger signal in direct DM searches.
Hence case 1) suffices in capturing the effect of the BMSSM operator in gaugino-like LSP scenarios where the scattering cross section on nucleons is minimal.   
We gather in Appendix~\ref{app:Zneu}, for both cases, all relevant expressions  to leading $\mathcal{O}(m_Z)$, including $\mathcal{O}(\eps_1)$ corrections. 

\subsection{Gaugino dark matter and direct detection}

Gaugino DM requires a $\mu$-parameter significantly larger than the lowest of $M_1$ and $M_2$. In this case, direct DM searches constitute a significant source of pressure on EW naturalness, which increases with the LSP mass~\cite{Perelstein1}. The tension stems from the fact that the tree-level fine-tuning is minimal for low $\mu$ values, while direct detection limits the Higgsino fraction of the LSP, thus favoring large $\mu$ values.  
A rather natural gaugino DM scenario could still be consistent with direct searches if the LSP is sufficiently light to avoid a significant decoupling of the Higgsino above the weak scale, $m_\chi\simeq10-30\,$GeV. Gaugino LSP's in this mass range are mostly bino-like in order to avoid excessively large chargino pair production cross-sections at LEP2. 
Note however that light bino DM thermal relics are typically overabundant due to their small hypercharge couplings to fermions, unless at least one of the following well-known exceptions~\cite{3exceptions} is realized. Bino annihilation into fermion pairs can be significantly enhanced through either $t-$channel exchange of light sfermions (mostly right-handed staus), or $Z$ or Higgs bosons resonances~\cite{DreesNojiri,resonant1,resonant2}, with $m_\chi\simeq m_{h,Z}/2$. Strong bino co-annihilation with either light right-handed staus with $m_{\tilde \tau_R}\simeq m_\chi$~\cite{staucoan1,staucoan2}, or light stops~\cite{BMSSMnir} are also possible. However, in the MSSM, a sufficient increase of the bino annihilation cross section through stau exchange or resonant enhancement is in conflict with collider constraints for $m_\chi\lesssim 15\,$GeV~\cite{LightBino1} and $m_\chi\lesssim 30\,$GeV~\cite{LightBino2}, respectively.

The SI scattering of $\chi$ onto nucleons is typically dominated by the SM-like Higgs $t-$channel exchange, whose relevant $h$-to-$\chi\chi$ coupling is
\be
\mathcal{L}_{h\chi\chi}=\frac{1}{2}g_{h\chi\chi}h\chi^T\chi\,,
\ee
with
\be\label{hXX}
g_{h\chi\chi}=g(\mathcal{N}_{\chi 2}-t_W\mathcal{N}_{\chi1})(\mathcal{N}_{\chi3}\sin\alpha+\mathcal{N}_{\chi4}\cos\alpha)+\delta g_{h\chi\chi}\,,
\ee
where the angle $\alpha$ parameterizes the orientation of the CP-even Higgs mass eigenstates relative to the vacuum (see Eq.~\eqref{cpevenstates}), and
\be\label{dgh}
\delta g_{h\chi\chi}=-2\sqrt{2}\frac{\eps_1v}{\mu}\left[2\cos(\alpha+\beta)\mathcal{N}_{\chi3}\mathcal{N}_{\chi4}+\cos\alpha\sin\beta\mathcal{N}_{\chi3}^2-\sin\alpha\cos\beta \mathcal{N}_{\chi4}^2\right]\,.
\ee
The effective Lagrangian in Eq.~\eqref{Lhino} modifies the Higgs coupling to $\chi$ pairs at $\mathcal{O}(\eps_1)$. These corrections arise on the one hand through modification of the $\mathcal{N}_{\chi i}$'s dictating the LSP composition as well as through the introduction of new Higgs-Higgsino interactions leading to Eq.~\eqref{dgh}. Note that contrary to the MSSM, $g_{h\chi\chi}$ no longer vanishes in the limit where $\chi$ is a pure Higgsino state ($\mathcal{N}_{k1,2}=0$), albeit the non-zero coupling only contributes at $\mathcal{O}(\eps_1^2)$ in scattering cross sections.\\   
Expanding to leading order in the $\tilde B/\tilde h$ mixing, the Higgs coupling to $\chi$ pairs in the bino-like LSP case is (see Appendix~\ref{app:Zneu}) 
\beq\label{hXXbino}
g_{h\chi\chi}^{\tilde B}\simeq \frac{2g'm_Zs_W}{\mu}\left(\frac{1}{\tan\beta}+\frac{M_1}{2\mu}-\frac{\eps_1 v^2}{\mu^2}\right)\,
\eeq
up to (neglected) $\mathcal{O}(\tan^{-2}\beta)$ and $\mathcal{O}(m_Z^2)$, where we assumed the decoupling limit in which $\sin\alpha\to -\cos\beta$, $\cos\alpha\to \sin\beta$. The leading term in Eq.~\eqref{hXXbino} is suppressed at large $\tan\beta$, and in this case the coupling is controlled by higher orders of $M_1/\mu$ growing with the LSP mass $m_ \chi\simeq M_1$. Equation~\eqref{hXXbino} shows that the effective operator in Eq.~\eqref{ops1} always increases the Higgs coupling to LSP pairs whenever used to make $m_h\gtrsim m_Z$ at tree-level, {\it i.e.} for $\eps_1<0$, unless $M_1$ and $\mu$ have opposite signs and $\mu\lesssim |M_1|\tan\beta/2$. Furthermore, the $\tan\beta$-suppression of the leading term in Eq.~\eqref{hXXbino} is typically much less effective in the BMSSM as $\tan\beta$ is limited by the condition~\eqref{tbmax}, while it can easily exceed $\mathcal{O}(10)$ in the MSSM. Therefore, at fixed DM mass and fine-tuning, the scattering cross section on nucleons is expected to be significantly larger in the BMSSM relative to the MSSM. We investigate in full numerical detail the BMSSM implications for the connection between fine-tuning and direct DM searches in the case of gaugino-like LSPs in Sec.~\ref{Gresults}.

\subsection{Relic density of Higgsino dark matter}

Higgsino DM, with a relatively light $\mu$-parameter, is typically more favored by EW naturalness, as apparent in Eq.~\eqref{FTindiv}. There is therefore no tension with direct searches in this case. The Higgs coupling to Higgsino-like LSP pairs in the limit of decoupled wino is (see Appendix~\ref{app:Zneu})
\beq\label{hXXhino}
g_{h\chi\chi}^{\tilde h-{\rm like}}\simeq  \frac{g' m_Zs_W}{2 M_1}
\left(1+\sin2\beta-\frac{\eps_1v^2}{\mu^2}\cos^22\beta\right)-\sqrt{2}\frac{\eps_1v}{\mu}\left(1-2\sin2\beta\right)\,,
\eeq
up to $\mathcal{O}(\mu/M_{1})$, assuming again the decoupling limit. The last term in Eq.~\eqref{hXXhino} originates from Eq.~\eqref{dgh}. Since $\cos2\beta<0$, the effective operator always reduces the MSSM-like contribution to the $h\chi\chi$ coupling whenever used to increase $m_h$ ($\eps_1<0$), while the direct BMSSM contribution $\delta g_{h\chi\chi}$ increases the overall coupling for $\tan\beta\gtrsim 3.7$. Note that the Higgs-to-LSP pair coupling remains sizable even in a limit where the gauginos are decoupled. \\

Other important quantities in the Higgsino-like LSP case are the mass splittings among the LSP, the next-to-lightest neutralino $\chi'$ and the lightest chargino $\chi^\pm$, which control the annihilation and co-annihilation processes that determine the relic density of Higgsino-like neutralinos. While in the renormalizable MSSM the four Higgsino states $\tilde h_{u,d}^0$,  $\tilde h_d^-$, $\tilde h_u^+$ are degenerate at tree-level, the effective operator of Eq.~\eqref{ops1} contributes to the lightest neutralino and chargino state mass splittings as (see Appendix~\ref{app:Zneu})
\beq
\delta m_\chi\equiv m_{\chi'}-m_{\chi}= -\frac{2\eps_1v^2}{\mu}+\frac{m_W^2}{M_2}
+\frac{m_Z^2s_W^2}{M_1}\,,
\eeq
\beq\label{charginosplit}
\delta m_{\tilde C}\equiv m_{\tilde C}-m_\chi = (1- \sin2\beta)\left(-\frac{\eps_1v^2}{\mu}+\frac{m_W^2}{2M_2}\right)+(1+\sin2\beta)\frac{m_Z^2s_W^2}{2M_1}\,,
\eeq
up to (neglected) terms of $\mathcal{O}(1/M_{1,2}^2)$. $\delta m_{\tilde C}$ and $\delta m_\chi$ as large as $\simeq34\,$GeV and $90\,$GeV, respectively, are obtained for $\mu=80\,$GeV, $m_A=300\,$GeV  and $\tan\beta=8$, in the $M_{1,2}\to \infty$ limit. These large $\delta m_{\tilde C}$ values allow for a scenario where the Higgsino-like LSP lies below the $W$ mass, thus strongly suppressing annihilation (as well as coannihilation) processes into weak gauge bosons in the early Universe which leads to the correct relic density while keeping the lightest chargino $\tilde C$ above the LEP2 kinematic limit~\cite{LEPbound}, see Eq.~\eqref{lepbound}.
We analyse in further detail the feasibility of such a scenario, with emphasis on its implications for EW naturalness, in Sec.~\ref{Hresults}.

\section{Dark matter implications for BMSSM naturalness} \label{sec:darkmatter}

We present the implications of DM constraints for BMSSM naturalness assuming all the DM consists of a lightest neutralino relic. We further assume that the LSP is exactly stable, {\it e.g.} protected by $R$-parity. We assume that all sfermions and the gluino are heavy enough to play a negligible role in the analysis. Although weak scale naturalness requires top (and left-handed bottom) squarks and to a (loop-factor) lower extent the gluino to be light~\cite{BG}, their presence can only qualitatively improve the model's agreement with DM direct detection data through large cancellations among {\it a priori} unrelated parameters in the low-energy theory. Unless it is possible to derive these relations from additional structures in specific UV completions, such cancellations should be interpreted as purely accidental, and as such they would always qualitatively worsen the overall degree of fine-tuning. Then, barring such accidents, the effect of light top squark in {\it e.g.} neutralino-nucleon scattering or neutralino annihilation only constitutes an $\mathcal{O}(1)$ correction to the processes considered in the present analysis. Hence, one is left with an irreducible source of pressure on naturalness through the $m_Z$ sensitivity in Eq.~\eqref{FTglobal} which, interestingly enough, is directly tied to DM observables.

Under this assumption, DM phenomenology and EW tree-level fine-tuning are described by only five parameters
\begin{equation}\label{parameters}
\tan\beta, \ m_A, \ \mu, \ M_1, \ M_2\,.
\end{equation}
We set the value of the BMSSM operator $\eps_1$ so that $m_h=125\,$GeV at tree-level, according to Eqs.~\eqref{Hmass} and~\eqref{dmh2eps}. We further assume a vanishing SUSY-breaking operator $\eps_2=0$. Non-vanishing values for the latter would affect our analysis as follows. $\eps_2>0$ would imply a smaller $|\eps_1|$ which in turn would reduce the BMSSM effects on the Higgsino sector, therefore loosening the connection between corrections to $m_h$ and DM observables inherent to the BMSSM. On the other hand, $\eps_2<0$ would push $|\eps_1|$ to unacceptably large values in order to maintain $m_h=125\,$GeV at tree-level. This would signal a breakdown of the EFT described in Sec.~\ref{sec:BMSSMreview}, and would therefore reintroduce fine-tuning through a large radiative correction to the Higgs mass. In both cases the situation would appear similar to that of the MSSM with a mostly radiatively induced Higgs mass with no relation to the neutralino sector.

Dark matter observables were computed numerically as follows. The neutralino-nucleon scattering cross sections for direct detection\footnote{We review the calculation of the SI  cross section, and specify our assumed values for the relevant hadronic form factors in Appendix \ref{app:SIscattering}.} and the thermal relic density were computed with {\tt micrOMEGAs 3.6.8}~\cite{MicrOmegas3}, where BMSSM Feynman rules were implemented with the help of the {\tt LanHEP} package~\cite{LanHEP}. In particular, we have taken special care to keep only effects to ${\cal{O}}(\eps_1)$ in $m_h$, according to Eq.~\eqref{dmh2eps}.
Whenever relevant, the Higgs boson width into neutralino pairs and electroweakino production cross sections at colliders were computed with the {\tt CalcHEP} package~\cite{CalcHEP04,CalcHEP12}. 

\subsection{Constraints}\label{sec:constraints}

We list in this section the DM related and collider contraints relevant to our analysis. These are:

\begin{itemize}
\item  Direct DM searches constraints from the first run of the LUX experiment~\cite{LUX}. This is the most stringent direct detection constraint to date in the mass range of interest, $m_\chi\gtrsim 15\,$GeV and below a few TeV. The current $90\%$ confidence level (CL) limit from LUX on the  (SI) DM-nucleon cross section peaks at  
\beq\label{LUXbound}
\sigma_{\rm SI}^{\rm LUX}\simeq7.6\times 10^{-46}\,{\rm cm}^2\,,
\eeq
for $m_\chi\simeq 33\,$GeV. The limit is significantly relaxed at larger masses, reaching {\it i.e.} $\sigma_{\rm SI}^{LUX}\simeq 1.1\times 10^{-44}\,$cm$^2$ for $m_{\chi}\simeq 1\,$TeV. We also occasionally use for illustration the projected sensitivities of future experiments with the XENON1T~\cite{XENON1T} and LZ ~\cite{LZ} detectors, which are expected to peak respectively at $\sigma_{\rm SI}^{\rm X1T}\simeq 2\times 10^{-47}\,$cm$^2$  for $m_\chi\simeq 55\,$GeV, and $\sigma_{\rm SI}^{\rm LZ}\simeq 1.4\times 10^{-48}\,$cm$^2$ for $m_\chi\simeq 55\,$GeV.
Whenever imposing this constraint, we further assume for simplicity that the local DM density has the canonical $0.3\,$GeV$\,$cm$^{-3}$ value~\footnote{This value, which is conventionally used by  direct DM searches experimental collaborations, is further supported by observations of galactic dynamics of the Milky Way~\cite{LocalDensity}.}, regardless of whether the  predicted DM density precisely coincides with the observed one. This is a reasonable approach for regions of parameter space which yield a relic density in the right ballpark, given that its computation by {\tt micrOMEGAs}  is only performed at tree-level, while sizable radiative corrections could arise in dominant annihilation channels~\cite{BjoernSUSYQCD,Baro:2009na,FawziOneLoop,AntonioEffVert}. 
\item The DM relic density derived from the combined (CMB+BAO+$H_0$) WMAP 9-year results~\cite{WMAP9}, which is
\beq\label{WMAP}
\Omega_{\rm CDM}^{\rm WMAP9} h^2 = 0.1153 \pm  0.0019\,,
\eeq
for its central value and standard deviation, respectively.
Again, due to theoretical uncertainties in the relic density calculation, we consider agreement with the WMAP result within three standard deviations as reasonably satisfactory.

\item The LEP bound on light chargino states~\cite{LEPbound}. The chargino mass constraint is only of crucial importance for light Higgsino DM with $m_\chi\lesssim 80\,$GeV. We require in our analysis the lightest chargino to be above the LEP2 kinematic limit 
\be\label{lepbound}
m_{\tilde C} \gtrsim 103\,{\rm GeV}\,. 
\ee
We have refrained from imposing the less stringent bound of $94\,$GeV~\cite{pdg} often adopted in the literature, which only applies to very specific configurations which are irrelevant here. Those include a largely destructive interference with a light sneutrino exchange~\cite{pdg} for chargino pair production at  $e^+e^-$ colliders. We also checked that Higgsino DM scenarios with $m_{\tilde C}<103\,$GeV yield chargino pair production cross sections always far above the ADLO combined limit at LEP2~\cite{LEPbound}.   
We will, however, allow for a $\simeq5\,$GeV loosening of Eq.~\eqref{lepbound}   when comparing to the tree-level Higgsino spectrum computed within {\tt micrOMEGAs}. This accounts for the maximal radiative corrections (dominated by stop and sbottom loops) to the neutral-charged Higgsino mass splitting allowed by EW precision data~\cite{GiudicePomarol}. Note that light charginos with a large Higgsino component  easily evade LHC constraints  due to their  small mass splitting with the LSP~\cite{Khachatryan:2014qwa,Aad:2014vma}. We have moreover explicitly checked that the production cross section for $\chi_2^0\chi_1^+$, falls below the sensitivity of ATLAS~\cite{Aad:2014nua}.

\item The LHC bound on the invisible decay branching ratio of the Higgs boson, which is relevant for bino-like DM of mass less than $m_h/2\simeq 63\,$GeV.
We adopt the $95\%$ CL limit resulting from a global fit to all existing LHC run 1 and LEP data from {\it e.g.} Ref.~\cite{Higgsatlast},   
\beq\label{BRinv}
{\rm Br}(h\to \rm inv)\lesssim 0.5\,.
\eeq
We recall that the above limit is not stricly limited to Higgs branching ratios into invisible particles but actually applies to the total Higgs branching ratio into all untagged final states, including {\it e.g.} jets. We conservatively assume here that Eq.~\eqref{BRinv} constrains the $h \to \chi\chi$ decay, whenever kinematically accessible.  

\item The stability condition of Eq.~\eqref{stab} for the EW vacuum in the presence of the effective BMSSM operator in Eq.~\eqref{ops1}, as well as $\tan\beta< |\eps_1^{-1}|$ in order to warrant good control of the Higgs boson mass within the EFT as discussed in Sec.~\ref{sec:BMSSMreview}.
\end{itemize}

\subsection{Direct detection of neutralino dark matter}\label{fullscan}

The SI scattering cross section of DM on protosn\footnote{The equivalent cross section on neutrons is of comparable magnitude unless DM interactions with quarks significantly violate weak isospin. The resulting cross section on large nuclei like Xenon could accidentally be significantly reduced if the DM couplings to protons and neutrons have a relative sign, see {\it e.g.} Ref.~\cite{isospinviolatingDM, ivdmMSSM}. Motivated by naturalness, we do not give in to this possibility in this paper and we consider direct searches on heavy nuclei as directly bounding the DM coupling to protons or, equivalently, neutrons.} is dominated by $t$-channel exchange of CP-even neutral Higgs bosons and resonant squark exchange. We only consider here the former contribution, since it directly relates to the EW fine-tuning defined in Eq.~\eqref{FTglobal}. {\it Barring accidental cancellations}, $s$-channel exchange of light squarks, albeit certainly of relevance for the third generation, would only increase the scattering cross section. In a limit where the SM-light Higgs boson exchange dominates ($m_A\gg m_h$), and assuming $m_\chi\gg m_p\simeq 0.93\,$GeV, the SI cross section  for DM  scattering on protons is approximately (see Appendix~\ref{app:SIscattering})
\beq\label{sigmaSIapprox}
\sigma_{\rm SI} \simeq  \sigma_{\rm SI}^{\rm LUX}\times \left(\frac{g_{h\chi\chi}}{0.036}\right)^2\,,
\eeq
where $g_{h\chi\chi}$ is the Higgs-to-neutralino pairs coupling defined in Eq.~\eqref{hXX} and $\sigma_{\rm SI}^{\rm LUX}$ is the best $90\%$ CL limit from the first LUX results, see Eq.~\eqref{LUXbound}. This relation illustrates the tension that exists in neutralino DM scenarios between current direct searches and a weak-size ($g\simeq 0.65$) DM coupling to SM fields, and the WIMP miracle to a broader extent. Equations~\eqref{sigmaSIapprox} and~\eqref{hXX} show that direct DM searches constrain the LSP composition to nearly pure current states.

\begin{figure}[!t]
\begin{center}
\hspace{-1.5cm}
\begin{tabular}{cc} 
\includegraphics[width=0.7\textwidth]{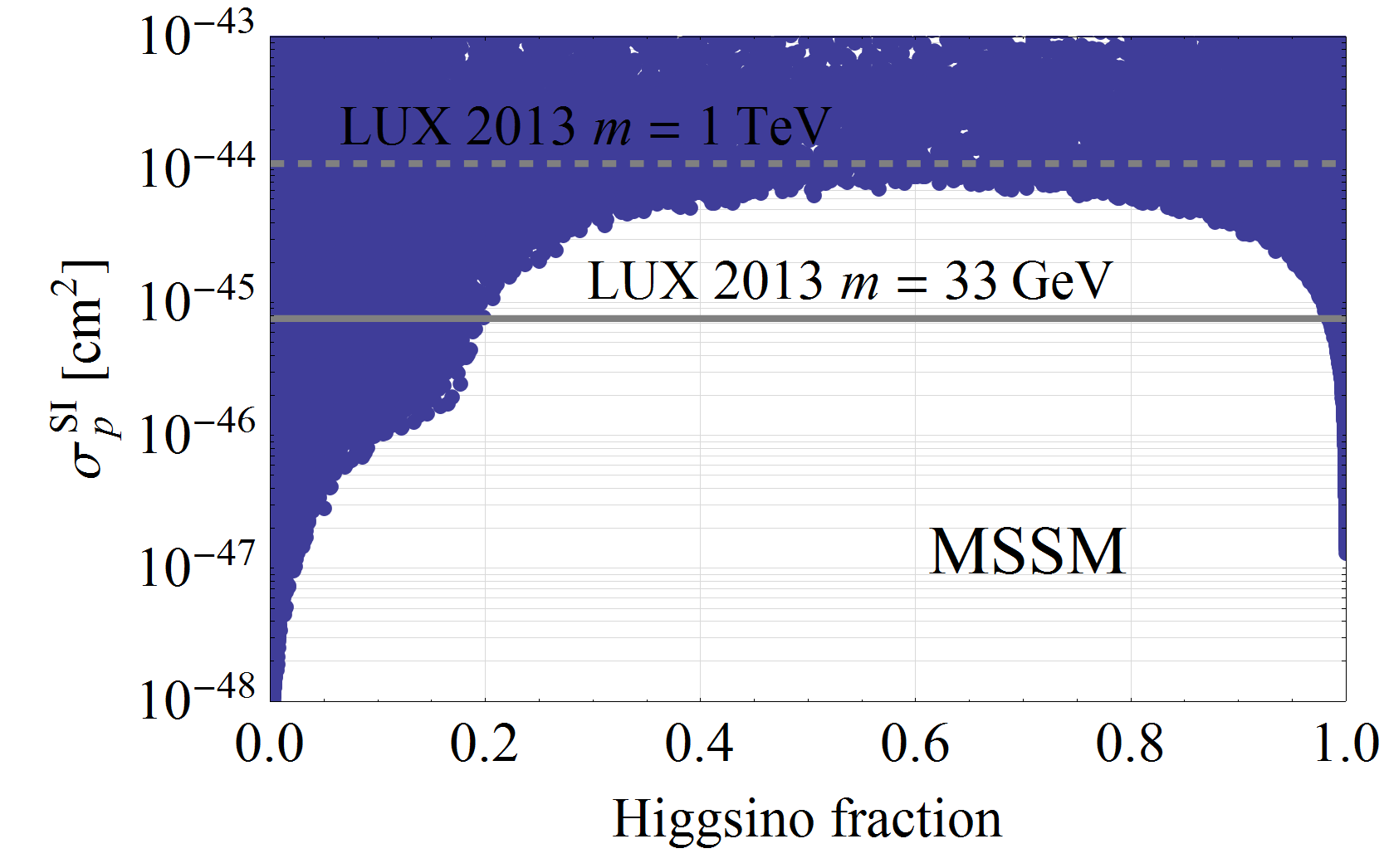}\\
\includegraphics[width=0.7\textwidth]{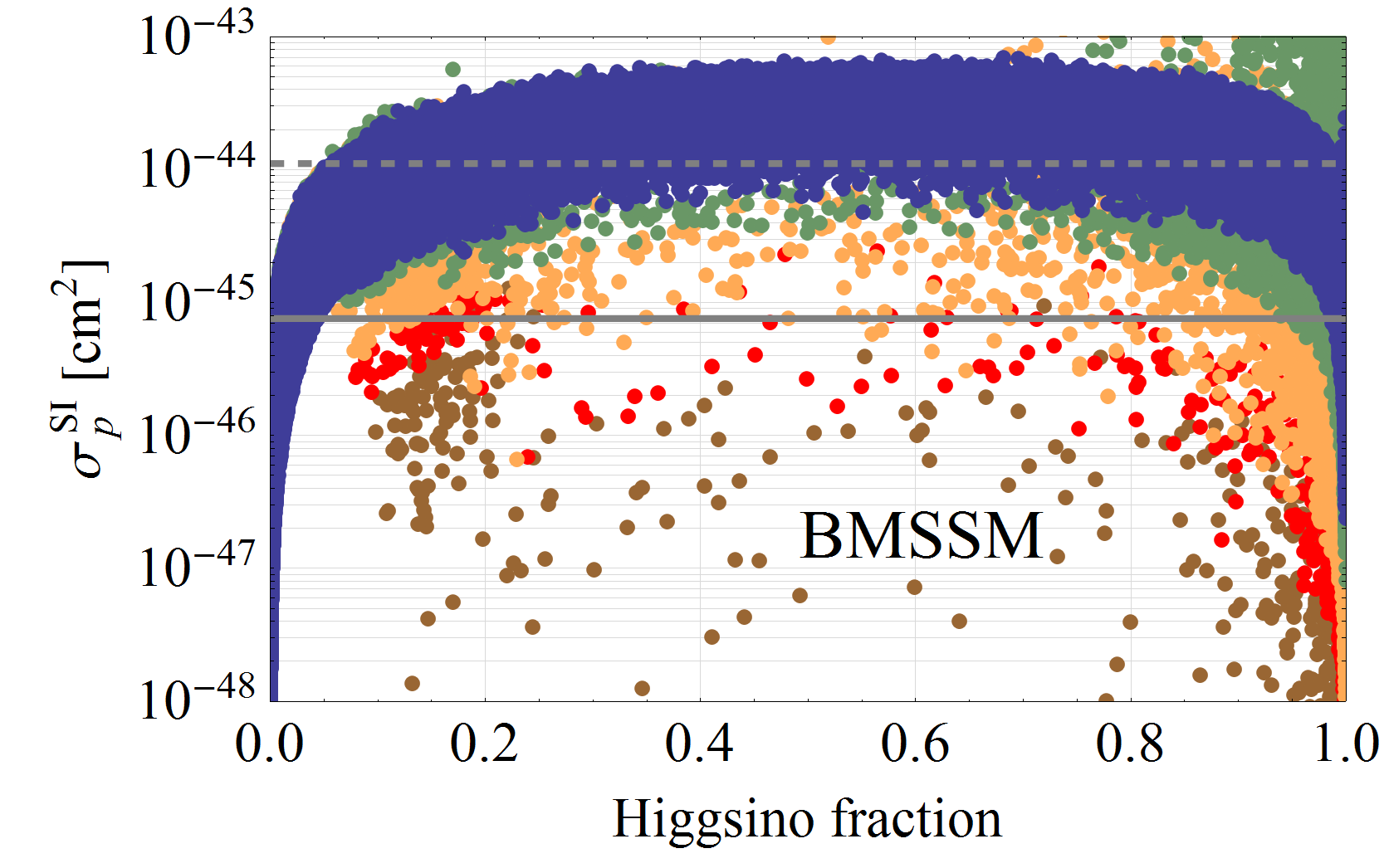}
\end{tabular}
\caption{Spin-independent cross section for DM scattering on protons as a function of the lightest neutralino Higgsino fraction in the MSSM (upper panel) and the BMSSM (lower panel).  The MSSM parameters are varied according to Eqs.~\eqref{fullscantb} and~\eqref{fullscan}. The solid (dashed) grey line shows the current $90\%$ CL limit from LUX~\cite{LUX} for $m_\chi=33\,$GeV ($1\,$TeV). In the BMSSM, colors correspond to different levels of log-sensitivity of the cross section with $\Delta_{\sigma_{\rm SI}}$ below 5 (blue), between 5 and 10 (green), 10 and 50 (orange), 50 and 100 (red) and above 100 (brown). For all points, the EW vacuum is exactly stable.}
\label{DD_compo}
\end{center}
\end{figure}

We performed a scan over the parameters in Eq.~\eqref{parameters} in order to quantitatively illustrate how present direct DM searches severely constrain the composition of neutralino DM to be close to pure gaugino or Higgsino states (thus suppressing $g_{h\chi\chi}$). Figure~\ref{DD_compo} shows the resulting distribution of $\sigma_{\rm SI}$ as a function of the Higgsino fraction $F_{\tilde H}\equiv\mathcal{N}_{\chi3}^2+\mathcal{N}_{\chi4}^2$, for $\sim\mathcal{O}(10^5)$ parameter space points both in the  MSSM limit ($\eps_1=0$) and the BMSSM case where $\eps_1$ was set to obtain $m_h=125\,$GeV at tree-level \footnote{In the MSSM, the observed value of the Higgs mass cannot be recovered classically. In this case stop quark parameters (among others) need to be  ajusted so that $m_h=125\,$GeV with the inclusion of radiative corrections. We assumed implicitly that this is the case, regardless of the amount of radiative fine-tuning $\Delta_{\rm rad}$ induced.}. 
We assumed $\tan\beta$ to be randomly distributed in the range 
\footnote{In this range both top and bottom Yukawa couplings in the MSSM remain  perturbative up to the GUT scale in models  with universal boundary conditions at the GUT scale~\cite{Djouadi:2005gj}.}

\beq\label{fullscantb}
{\rm MSSM}\ :\ \tan\beta \in [2,50]\,,\quad {\rm BMSSM}\ : \ \tan\beta\in [2,10]\,,
\eeq  
while the remaining four parameters were randomly varied, with uniform logarithmic distributions, within the ranges\footnote{The higher end of the considered intervals for $m_A$ and $M_{1,2}$ implies SUSY breaking soft terms potentially much larger than the BMSSM scale $M$, which would invalidate the supersymmetric EFT approach employed in our analysis. A possible workaround is to assume that the new interactions beyond the MSSM are relatively strong, {\it i.e.} $\lambda_1\sim 4\pi$. In this case, for $|\eps_1|\simeq 0.1$ and $\mu\simeq 100\,$GeV, the EFT cut-off could be raised to $M\sim\mathcal{O}(10\,$TeV). }
\beq\label{fullscan}
m_A \in [150, 8000]\, {\rm GeV}\,, \quad \mu  \in [55, 8000]\, {\rm GeV}\,, \\ 
M_1\in [10,8000]\, {\rm GeV}\,, \quad
M_2\in [100,8000]\, {\rm GeV}\,,
\eeq
in both the MSSM and BMSSM cases. Note that either of $M_1$ and $M_2$ could in principle carry a relative sign with respect to $\mu$. In a basis where $\mu>0$, negative $M_{1,2}$ values could yield large cancellations in the Higgs exchange amplitude for direct detection (see for instance Eq.~\eqref{hXXbino}), which would lead to SI cross sections orders of magnitude below the present LUX limit~\cite{Perelstein1,Perelstein2,LindnerNDM,DMblindspots}.  Given the {\it a priori} accidental nature of such cancellations, we do not consider these blind spots as natural regions of the parameter space. Away from these regions, the relative signs between $\mu$ and $M_{1,2}$ do not yield significantly different predictions for SI scattering or annihilation cross sections. Hence, we  choose to focus on positive values only. Unless specified otherwise, we discard points which do not satisfy the kinematic LEP2 bound on the lightest chargino of Eq.~\eqref{lepbound}, as well as those in the BMSSM for which the vacuum is not stable according to Eq.~\eqref{stab} or where $|\epsilon_1|\tan\beta\leq 1$.

The upper panel of Fig.~\ref{DD_compo} shows that light LSP's with $m_\chi\sim\mathcal{O}(100\,$GeV) are in at least one order of magnitude tension with the LUX experiment  in the MSSM, unless they are close to pure gaugino  ($F_{\tilde H}\lesssim0.2$) or pure Higgsino  ($F_{\tilde H}\gtrsim 0.98)$ states~\footnote{A qualitatively similar result was obtained in Ref.~\cite{Perelstein1}. Our results differ to some extent quantitatively mostly due to the use of more up-to-date nuclear form factors as decribed in Appendix~\ref{app:SIscattering}.}. Heavy LSP's around $m_\chi\sim\mathcal{O}(1\,$TeV) could  however be consistent with LUX regardless of their composition, but at the price of significant fine-tuning. The lower panel of Fig.~\ref{DD_compo} demonstrates that these results persist qualitatively in the BMSSM, provided no large cancellations occur in the cross section. We point out that all sets of parameters with $F_{\tilde H}\simeq 0.5$ and $\sigma_{\rm SI}$ orders of magnitude below the current best LUX limit  rely on accidental cancellations and thus display a significant sensitivity to a small variation of the MSSM parameters. Strong cancellations in the BMSSM can occur between the up- and down-type quark contributions to the scattering amplitude on protons when this is dominated by light Higgs exchange. This cancellation arises in regions of parameter space with $\alpha>0$, which, as explained in Appendix~\ref{app:SIscattering}, is genuine to the BMSSM. Inspired by the fine-tuning measure associated with the $m_Z$ sensitivity in Eq.~\eqref{FTglobal}, we use a logarithmic measure to quantify the sensitivity of the SI scattering cross section
\beq
\Delta \sigma_{\rm SI} \equiv \sqrt{\sum_{p} \left(\frac{d\log\sigma_{\rm SI}}{d\log p}\right)^2}\,,
\eeq
with $p=\mu,M_1,M_2,m_A,\tan\beta$. Figure~\ref{DD_compo} shows that light LSP's with $\Delta\sigma_{\rm SI}\lesssim10$ only agree with the LUX results for either $F_{\tilde H}\lesssim0.1$ or $F_{\tilde H}\gtrsim 0.95$. We further analyse in greater detail in the next subsections the impact of DM constraints on the BMSSM fine-tuning, as well as the corresponding differences with respect to the renormalizable MSSM, for gaugino-like and Higgsino-like LSP. 

\subsection{Gaugino dark matter}\label{Gresults}
We consider here gaugino-like DM scenarios, which occur when $M_1$ and/or $M_2$ are much smaller than $\mu$. In this section we focus on direct detection signals as in these scenarios they alone already significantly constrain EW naturalness. Further demanding the observed DM relic density to be thermally generated would require  specific adjustements of unrelated parameters, which could be interpreted as an extra source of fine-tuning. Given the different origin of the latter, we do not attempt to combine it with the weak scale sensitivity in Eq.~\eqref{FTglobal}, which we thus regard as a lower bound on the overall fine-tuning of the model. For concreteness, we focus on sets of parameters where the LSP projection on Higgsino states is $F_{\tilde H}<0.3$. 
\begin{figure}
\begin{center}
\includegraphics[width=0.7\textwidth]{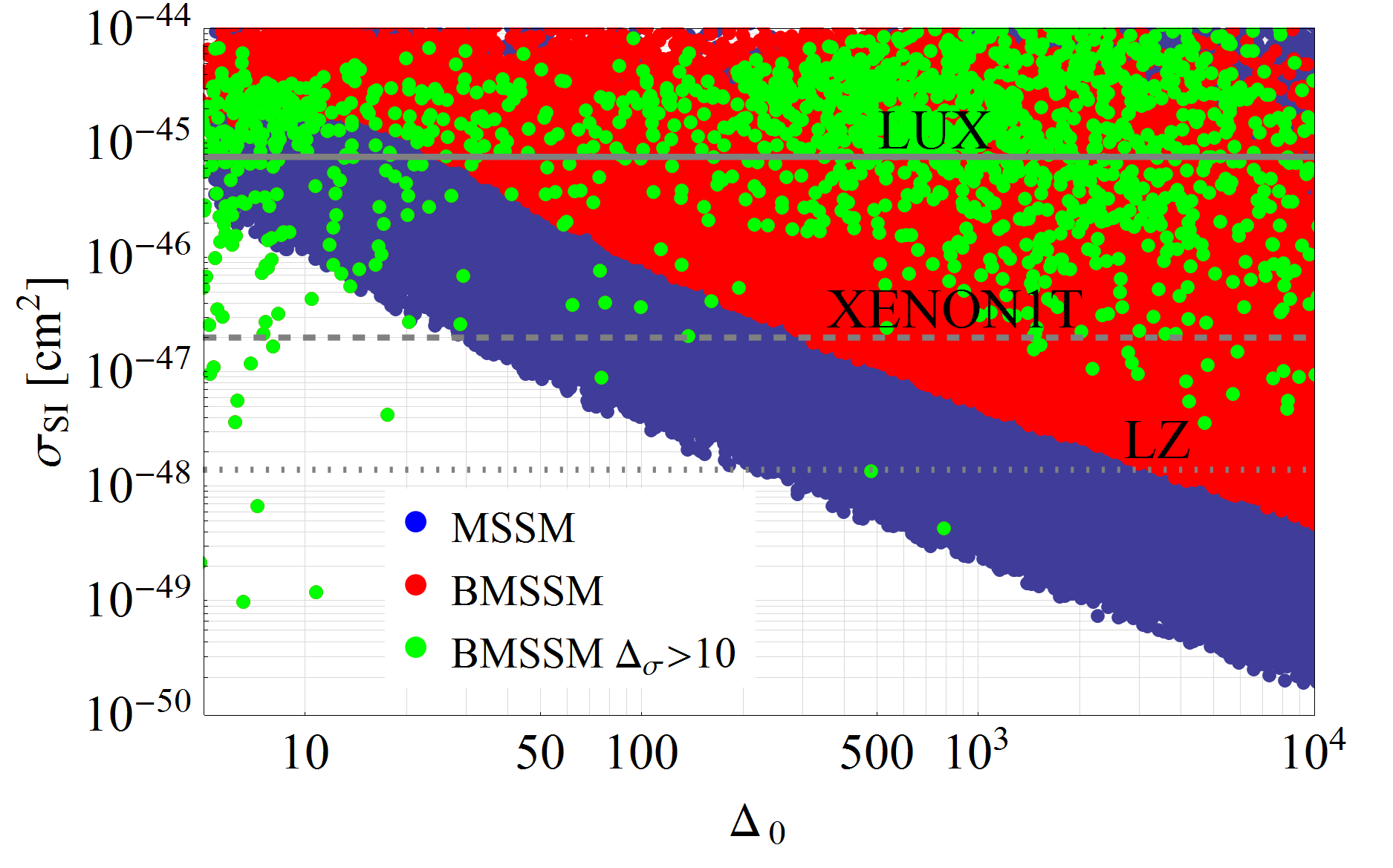}
\caption{Spin-independent DM scattering cross section off protons as a function of EW fine-tuning for gaugino DM with $F_{\tilde H}<0.3$. Green points denote sets of parameters in the BMSSM that lead to accidentally small scattering cross section with $\Delta\sigma_{\rm SI}>10$. The solid, dashed and dotted line represent the best sensitivities of the LUX, XENON1T and LZ experiments, respectively. For all points the EW vacuum is exactly stable.}
\label{sigma_FT_Gino}
\end{center}
\end{figure}

\begin{figure}[!t]
\begin{center}
\begin{tabular}{cc}
\includegraphics[width=0.8\textwidth]{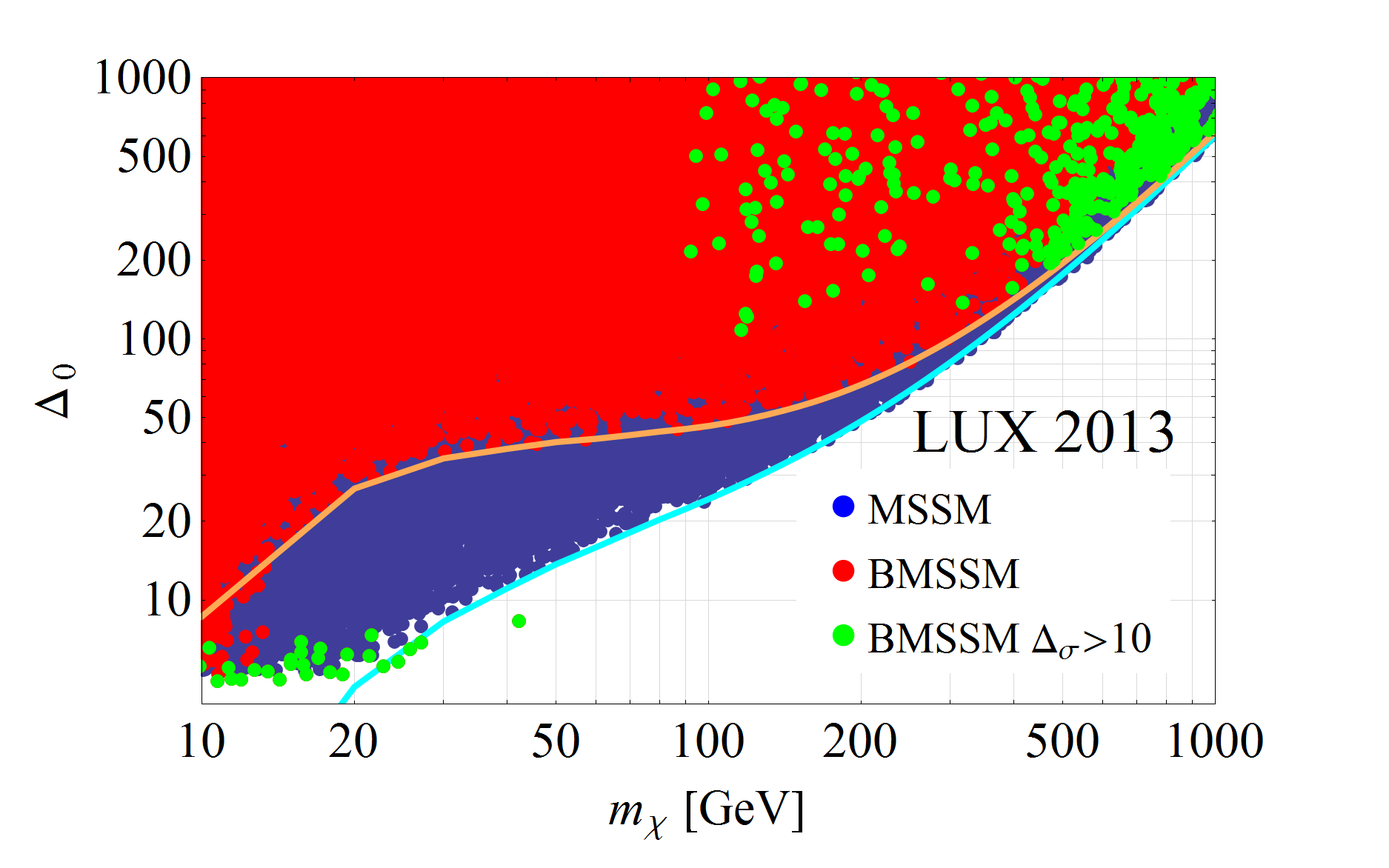}\\
\hspace{0.5cm}\includegraphics[width=0.8\textwidth]{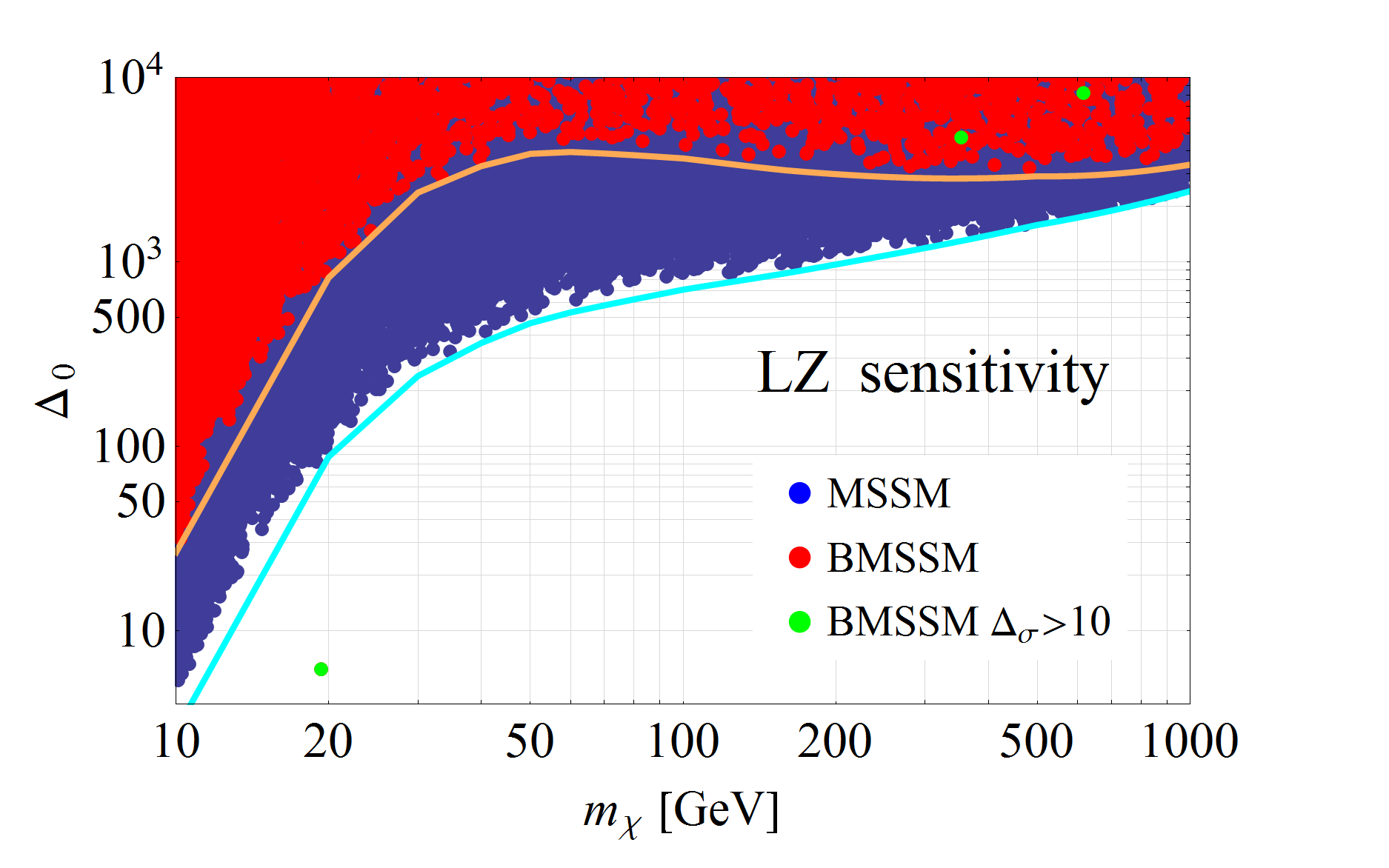}
\end{tabular}
\caption{EW fine-tuning as a function of the lightest neutralino mass for gaugino DM ($F_{\tilde H}<0.3$), imposing the current LUX limit (upper panel) or the projected LZ sensitivity (lower panel). The low fine-tuning BMSSM points in green arise at the expense of a significant accidental cancellation in the scattering cross section of $\Delta \sigma_{\rm SI}>10$. The cyan (orange) line denotes the minimal fine-tuning in the MSSM (BMSSM) derived through the approximate analytical diagonalization of the neutralino mass matrix, as shown in Appendix~\ref{app:Zneu}. For all points the EW vacuum is exactly stable.}
\label{fig:deltamchiBino}
\end{center}
\end{figure}

We plot in Fig.~\ref{sigma_FT_Gino} their corresponding distributions in the $\sigma_{\rm SI}-\Delta_0$ plane for both the MSSM and BMSSM. We observe that the minimal scattering cross section in the BMSSM is a factor of $\simeq 25$ larger than in the MSSM for a given fine-tuning level and for roughly all values of $\Delta_0$. This difference is a direct consequence of the small $\tan\beta$ requirement of Eq.~\eqref{tbmax} which warrants the consistency of the effective BMSSM approach. The minimal  cross section in both cases is dominated by the $t$-channel exchange of the SM-like Higgs, whose coupling to DM pairs is approximately $\propto \tan^{-1}\beta +m_\chi/(2\mu)-\eps_1v^2/\mu^2$, as shown in Eq.~\eqref{hXXbino} for bino DM. For $m_\chi\lesssim 2\mu/\tan\beta$, the first term dominates and the BMSSM minimal cross section is larger by roughly the square of the ratio of the MSSM maximal $\tan\beta$ to the BMSSM one, which is assumed here to be around 5. For larger LSP masses, the dominant contribution to the Higgs-to-DM pairs coupling is $\propto m_\chi/(2\mu)$ which is similar in both models. Note that larger $m_\chi$ in a gaugino DM scenario implies larger fine-tuning for fixed $F_{\tilde H}$. We show in Fig.~\ref{fig:deltamchiBino} the minimal fine-tuning achievable in the MSSM and BMSSM for a given DM mass under the current LUX constraint and that of a future LZ experiment. In the mass region where direct searches are most sensitive, $m_\chi\simeq 30-50\,$GeV, the LUX experiment forces the BMSSM, barring accidental cancellations (or equivalently for $\Delta\sigma_{\rm SI}<10$), to be at least a factor of $\simeq 4$ more fine-tuned than the MSSM, again due to the low $\tan\beta$ restriction. In the same LSP mass region, this situation will be further aggravated to a point where the BMSSM will be a factor  $\simeq 10$ more fine-tuned than the MSSM if no WIMP DM is observed at the future LZ experiment. We also note that a non-negligible fraction of the scenarios with $m\chi\lesssim\mathcal{O}(10)\,{\rm GeV}$ evading direct detection constraints are in tension with the invisible Higgs decay constraint of Eq.~\eqref{BRinv}. These points are in any case difficult to reconcile with the relic density constraint~\cite{LightBino1,LightBino2}. For $m_\chi\gtrsim200\,$GeV the BMSSM and MSSM minimal levels of fine-tuning imposed by LUX are comparable and worse than a few percent.     
The solid lines on Fig.~\ref{fig:deltamchiBino} denote the approximate minimal fine-tuning in agreement with LUX and LZ sensitivities derived through an analytical diagonalization of the neutralino mass matrix, as detailed in Appendix~\ref{app:Zneu}. 

\subsection{Higgsino dark matter}\label{Hresults}
We now move to consider Higgsino DM, which corresponds to $\mu$ being much smaller than $M_{1,2}$. We focus for concreteness on sets of parameters where the LSP has less than $30\%$ projection on gaugino states, {\it i.e.} $F_{\tilde H}>0.7$. 
Improvement of direct searches for Higgsino DM does not exert immediate pressure on naturalness as it would only force further decoupling of the gauginos, which does not reintroduce fine-tuning until $M_{1,2}$ enter the multi-TeV range. However, a tighter connection between DM and naturalness arises from imposing the thermal relic density constraint. Albeit favored by naturalness,  Higgsino LSP is typically not the most favorable DM candidate since it annihilates too efficiently into weak bosons in the early Universe, unless the DM is sufficiently heavy, $m_\chi\simeq\mu\sim\mathcal{O}(1\,$TeV), which in turn reintroduces large fine-tuning. This is a well-known result in the MSSM~\cite{welltempered}. We show below that this conclusion still holds in the BMSSM. A possible way-out is to make the LSP light enough so that the (co-)annihilation channels are kinematically closed. This happens for $m_\chi\lesssim m_W-T_f$, where $T_f\simeq m_\chi/20\simeq 3-5\,$GeV is the typical thermal DM energy at freeze-out. In the MSSM, however, this would lead to a light chargino below the $W$ mass, which is excluded by direct LEP searches (see Eq.~\eqref{lepbound}). 
The BMSSM operator is crucial in relaxing this tension due to a potentially significant contribution to the chargino/LSP mass splitting at $\mathcal{O}(\eps_1)$~\cite{BMSSMedsjo}. We show below that such a light Higgsino DM scenario is marginally resurrected in the BMSSM, at the expense of a one part in ten sensitivity to small variations of the model's parameters~\footnote{A much more optimistic result was obtained in Ref.~\cite{BMSSMedsjo}, where a looser chargino mass bound of $m_{\tilde C}\gtrsim 94\,$GeV was assumed. We numerically checked that all scenarios with light Higgsino LSP of right abundance and a chargino below the LEP2 kinematical limit are in at least a factor of few tension with the combined LEP2 constraint on the chargino pair production cross section at $e^+e^-$ colliders~\cite{LEPbound}.} .
\begin{figure}[!t]
\begin{center}
\begin{tabular}{cc}
\includegraphics[width=0.8\textwidth]{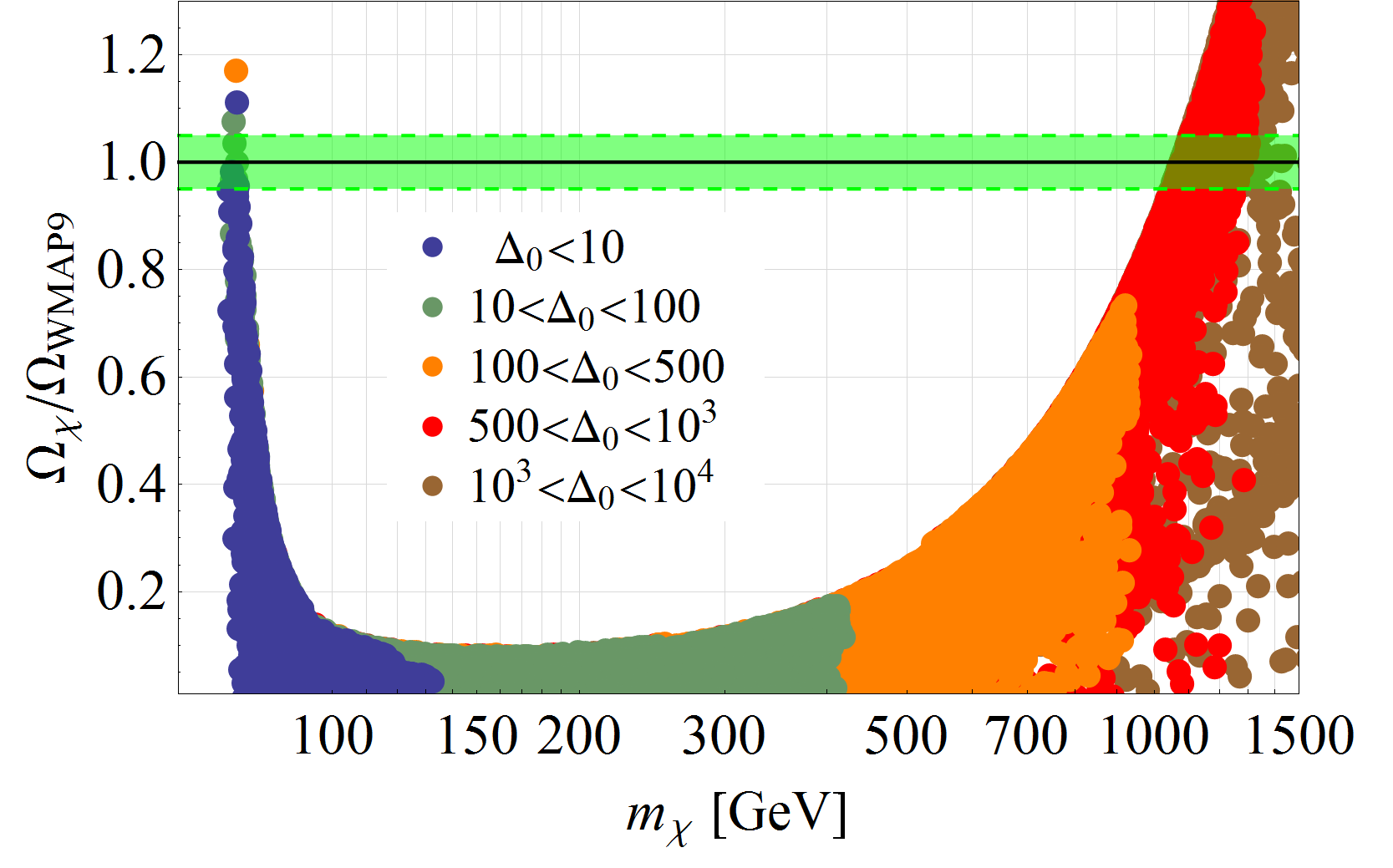}\\
\hspace{-0.6cm}\includegraphics[width=0.84\textwidth]{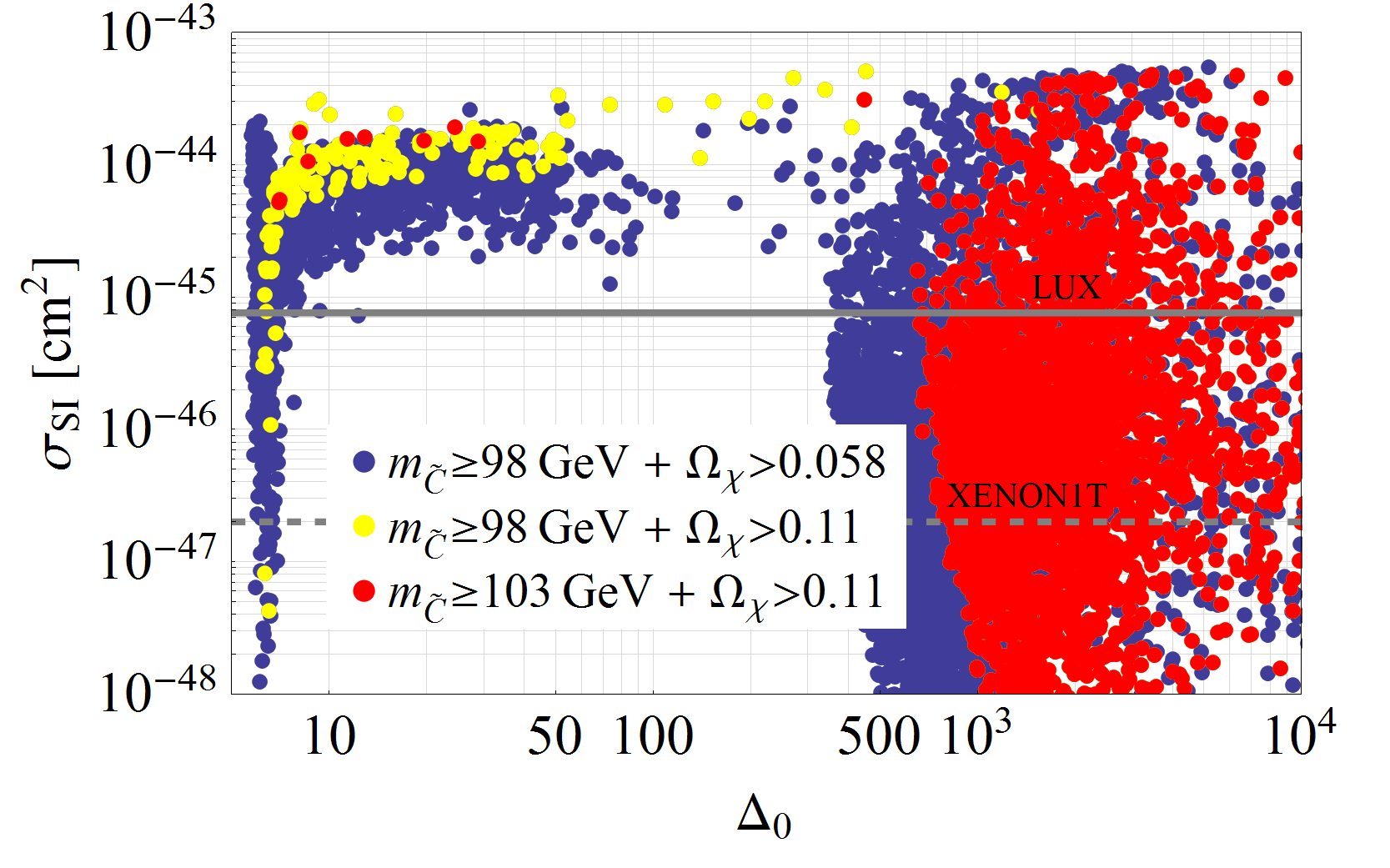}
\end{tabular}
\caption{[Upper panel] Neutralino relic density as a function of the LSP mass for Higgsino DM. The light green band depicts the  $3\sigma$-range favored by WMAP-9 for cold DM density values. Colors denote the variation of the EW fine-tuning $\Delta_0$ defined in Eq.~\eqref{FTglobal} with $m_\chi$. For $m_\chi\gtrsim 150\,$GeV, $\Delta_0\simeq \Delta_\mu \simeq \mathcal{O}(10)\times (m_\chi/150\,{\rm GeV})^2$. [Lower panel] Spin-independent scattering cross section as a function of EW fine-tuning $\Delta_0$ under the relic density constraint for Higgsino DM. Colors correspond to different requirements on the chargino mass and minimal amount of relic density.}
\label{fig:HiggsinosRelic}
\end{center}
\end{figure}

We show in the upper panel of Fig.~\ref{fig:HiggsinosRelic} the relic density predicted in the BMSSM as a function of the DM mass, together with contours of tree-level EW fine-tuning $\Delta_0$. The LSP relic density reaches the WMAP9 level of Eq.~\eqref{WMAP} for $m_\chi\lesssim 80\,$GeV and $m_\chi\gtrsim 1.1\,$TeV. 
As argued above, for Higgsino LSP above the $W$ mass the resulting DM energy-density at freeze-out is overly suppressed, due to (co-)annihilations into gauge bosons, unless  $m_\chi\gtrsim \mathcal{O}(1\,$TeV). The extreme efficiency of these channels results from the near mass degeneracy of the LSP and the other Higgsino states, as well as the sizable $SU(2)$ coupling among Higgsinos. As shown in Fig.~\ref{fig:HiggsinosRelic} this conclusion barely changes in the presence of the BMSSM operator. The latter lifts the tree-level degeneracy among Higgsino states, which in turn suppresses co-annihilation processes, and modifies their coupling to the $W$ and the $Z$ at $\mathcal{O}(\eps_1)$. However, both effects scale as $\propto|\eps_1|v^2/\mu^2$ and are suppressed down to negligible levels for $m_\chi\simeq1\,$TeV, yielding a fine-tuning of a permil or worse, comparable to the MSSM. 

\begin{figure}[!t]
\begin{center}
\begin{tabular}{cc}
\includegraphics[width=0.8\textwidth]{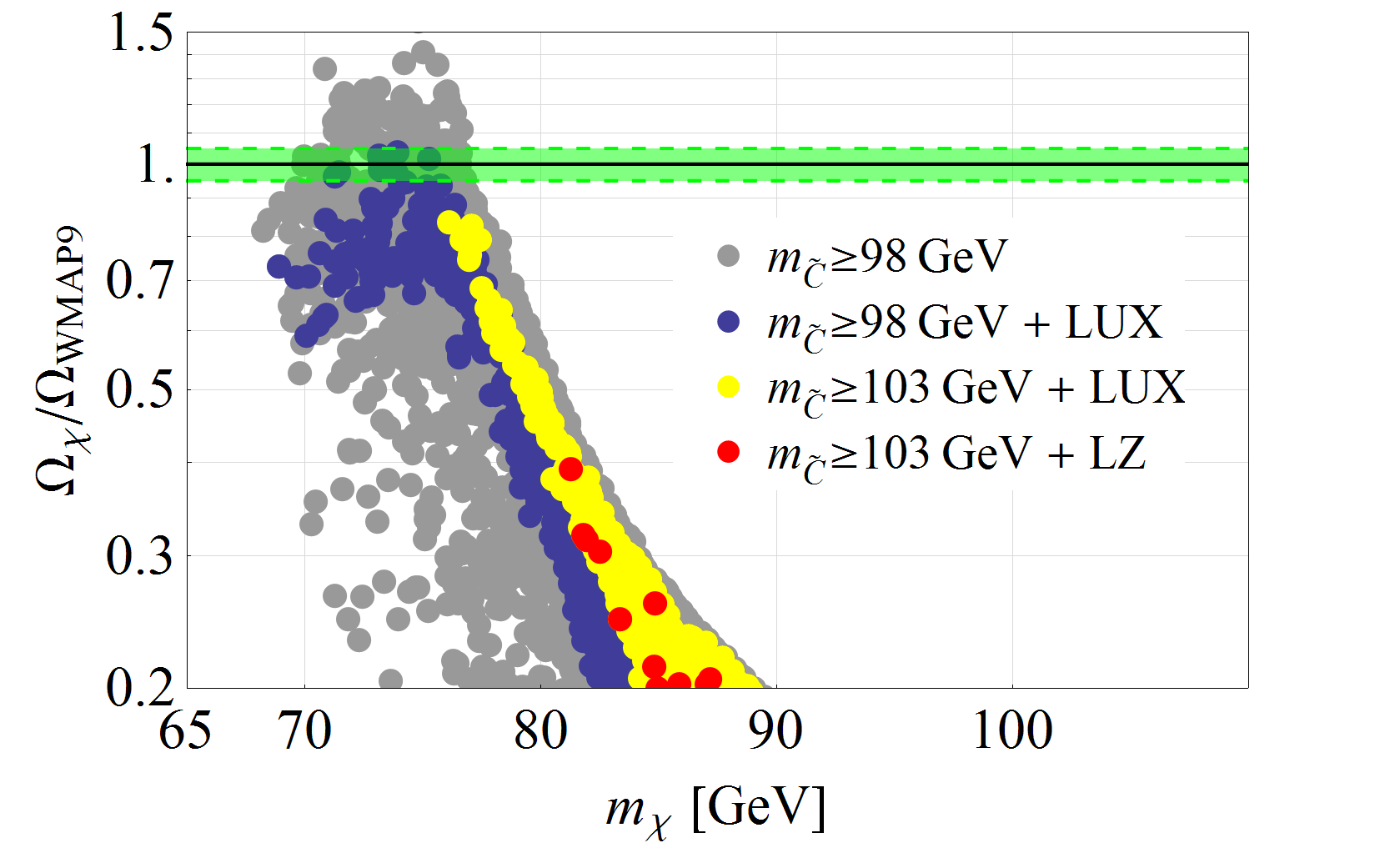}\\
\includegraphics[width=0.8\textwidth]{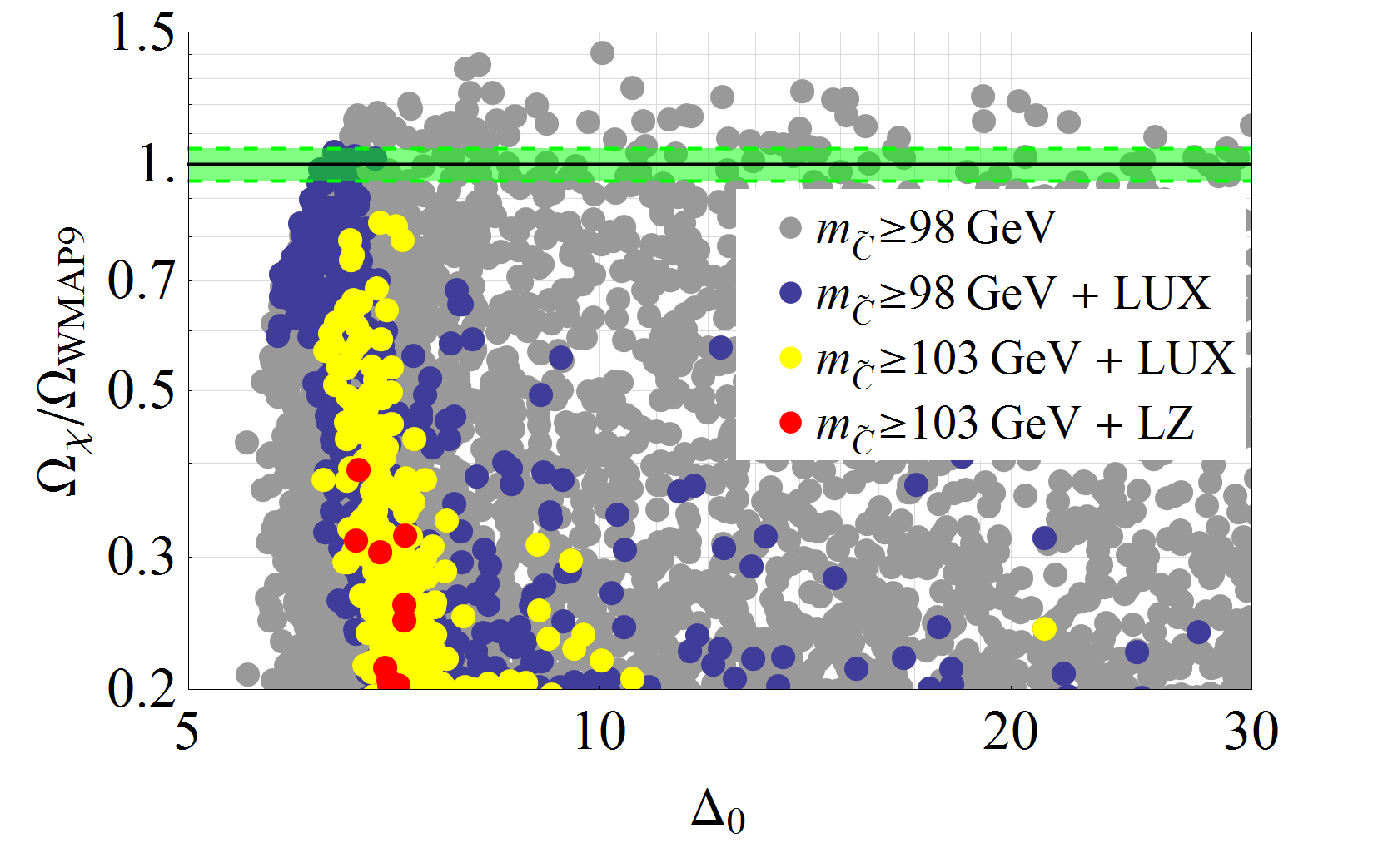}
\end{tabular}
\caption{Neutralino relic density in the BMSSM as a function of the LSP mass (upper panel) and the EW fine-tuning $\Delta_0$ (lower panel) for Higgsino-like LSPs below the threshold of EW boson pair production. Colors denote the requirement to satisfy various constraints on the chargino mass and the SI DM scattering cross section probed by direct searches as explained in Sec.~\ref{sec:constraints}.}
\label{fig:HiggsinosZoom}
\end{center}
\end{figure}

\begin{figure}[!t]
\begin{center}
\includegraphics[width=0.8\textwidth]{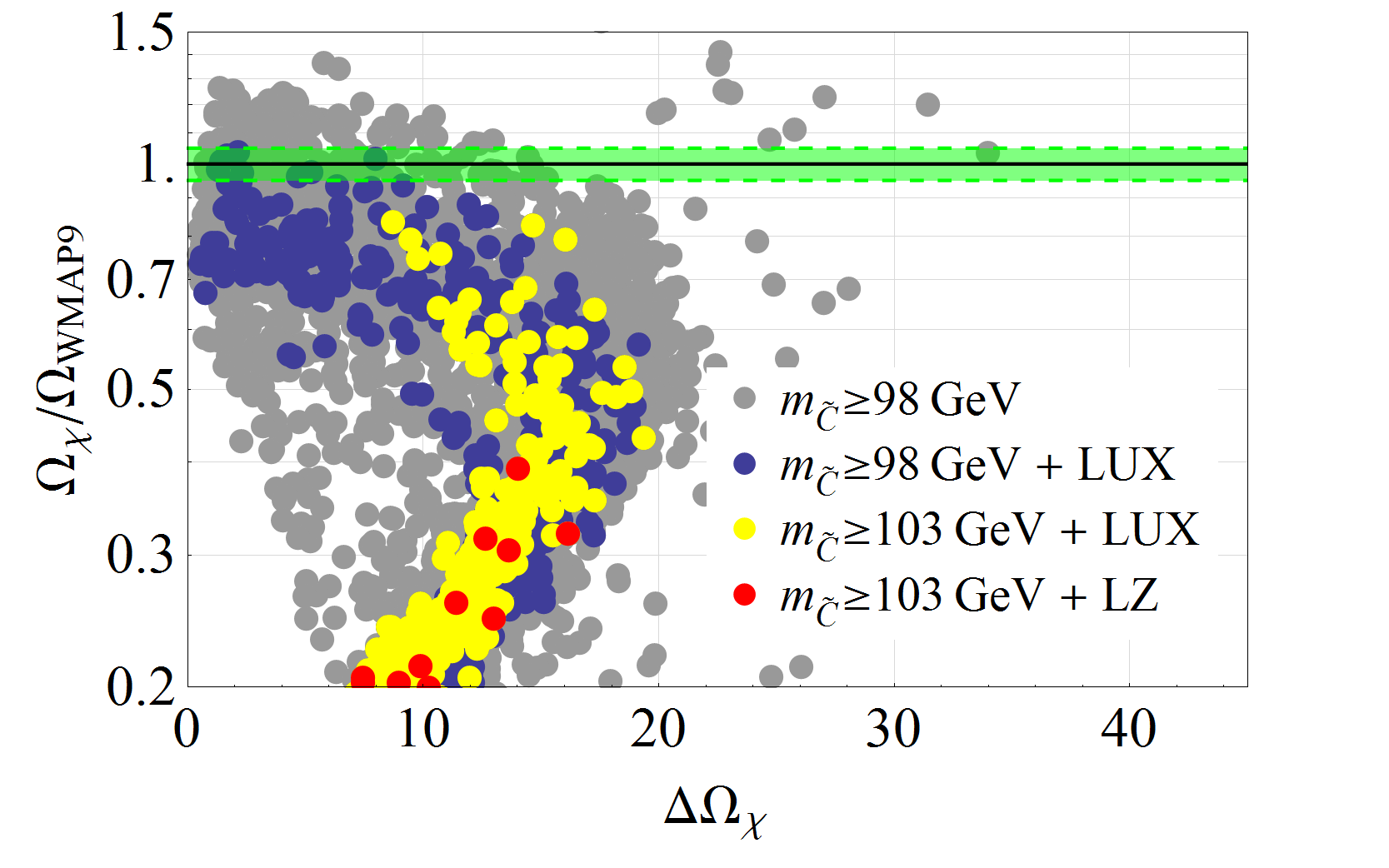}
\caption{Relic density of Higgsino DM as a function of its log-sensitivity to fundamental parameters as predicted in the BMSSM for $m_\chi\lesssim m_Z$. The light green band depicts the  $3\sigma$-range favored by WMAP-9 for cold DM density values. Colors denote the requirement to satisfy various constraints on the chargino mass and the SI DM scattering cross section probed by direct searches as explained in Sec.~\ref{sec:constraints}.}
\label{fig:HiggsinosAccidental}
\end{center}
\end{figure}

The light Higgsino region below $m_W$ is genuine to the BMSSM. With a moderate fine-tuning better than ten percent, as shown in the lower panel of Fig.~\ref{fig:HiggsinosZoom}, this region holds promise of being the only possible island of naturalness for Higgsino DM. Yet, a few comments are in order. 
\begin{itemize}
\item $\eps_1$ values as negative as $\simeq-0.12$ are required in order to maximize the mass splitting with the lightest chargino. Such large values are only  attainable for $\tan\beta\simeq 8-10$, which nearly saturates the upper bound of Eq.~\eqref{tbmax} and thus corresponds to a regime where $m_h$ starts being sensitive to (neglected) higher orders in inverse power of the cutoff scale $M$. Similarly large $\eps_1$ values could however be obtained at smaller $\tan\beta$ if some contribution of the SUSY-breaking operator in Eq.~\eqref{ops2} is introduced with $\eps_2\lesssim0$. But in this case the connection between the Higgs mass and the DM phenomenology is partially lost.

\item A not-too-heavy wino of $\mathcal{O}($few 100) GeV should be present in the spectrum in order to yield the necessary extra $\mathcal{O}($few GeV) contribution to the LSP/chargino mass splitting. The presence of the wino not far above the LSP mass would however induce a significant wino component of the LSP, which is constrained by direct detection.
The upper panel of Fig.~\ref{fig:HiggsinosRelic} illustrates the impact of null results at the current LUX and the future LZ experiments on the light Higgsino DM scenario. The scenario is marginally consistent with the current LUX limits. A stronger direct detection constraint would push the wino to higher masses, which forces the LSP mass to increase through a reduced mass splitting with the chargino. Furthermore, once $m_\chi\gtrsim75\,$GeV, LSP (co-)annihilation processes through off-shell weak gauge bosons become efficient in depleting the relic abundance. For instance, a null result at the LZ detector would then imply that light Higgsino DM in the BMSSM cannot form more than $\mathcal{O}(50\%)$ of the observed DM abundance, if thermally produced in the early Universe. A simple inspection of the lower panel of Figs.~\ref{fig:HiggsinosRelic} and~\ref{fig:HiggsinosZoom} leads to the same conclusion.

\item As shown in {\it e.g.} Fig.~\ref{fig:HiggsinosZoom}, current direct searches and collider bounds limit the DM abundance to $\simeq 80\%$ of the observed value. This result assumes that the mass splitting of the LSP with the chargino is given by the tree-level relation of Eq.~\eqref{charginosplit}. Sizeable radiative corrections to the Higgsino mass splittings  can arise if the mixing between stop quarks is large~\cite{GiudicePomarol}. The correction cannot exceed $\simeq 5\,$GeV without inducing an overly large contribution to the so-called $\rho$-parameter relative to the SM~\cite{GiudicePomarol}. We show in  Figs.~\ref{fig:HiggsinosRelic} (lower panel) and~\ref{fig:HiggsinosZoom} that under the assumption of a supplementary $\simeq 5\,$GeV radiative contribution to $\delta m_{\tilde C}$ all of the observed DM could consist of a light Higgsino LSP and satisfy current limits. Any slight improvement of either the chargino  bound at the LHC or the SI cross section at forthcoming direct detection experiments would strongly disfavor this scenario.   

\item The narrowness of the LSP mass region suggests a non-negligible sensitivity of the relic density prediction to the model's parameters, in particular $\mu$ which dominantly controls the LSP mass. In order to better quantify the latter we used the logarithmic measure 
\beq
\Delta_{\Omega_\chi} \equiv \sqrt{\sum_p \left(\frac{d\log\Omega_\chi}{d\log p}\right)^2}\,,
\eeq
with $p$ running over $\mu,M_1,M_2,m_A,\tan\beta$. Figure~\ref{fig:HiggsinosAccidental} shows that this sensitivity does not exceed $5\%$ for $m_\chi\lesssim 90\,$GeV. Albeit not completely free of fine-tuning, the light Higgsino DM in the BMSSM still appears qualitatively more natural than its $\mathcal{O}($TeV) counterpart in the MSSM.
\end{itemize} 
To summarize, the BMSSM scenario with a light Higgsino DM below the weak gauge boson threshold, albeit displaying a low EW fine-tuning, does lean on specific assumptions among unrelated parameters. This signals an additional sensitivity to the model's parameters, besides  the measure of Eq.~\eqref{FTglobal}, which we estimate to be at least of one part in ten. Moreover, this light Higgsino LSP scenario is probably subject to a mild radiative fine-tuning imposed by direct stop searches at the LHC. For $m_\chi\simeq 80\,$GeV, the current lower bound on the lightest stop mass is $\simeq 650\,$GeV~\cite{Chatrchyan:2013xna,Aad:2014kra}, unless the stop lies in the stealth region~\cite{stealthstop}, which roughly corresponds to an $\simeq \mathcal{O}(10\%)$ radiative fine-tuning. Once the gauge boson channel opens up, its efficiency in depleting DM pushes EW fine-tuning both in the MSSM and the BMSSM in the permil territory if neutralino LSPs are to constitute all of DM in the universe.  
\section{Conclusions} \label{sec:outro}

We considered SUSY extensions beyond the MSSM where the characteristic scale of the new sector is parametrically larger than that of the minimal SUSY spectrum. This separation of scales allows for an EFT description of the new dynamics in terms of MSSM superfields and symmetries. There is a unique higher dimensional operator at lowest order involving only Higgs fields, which easily raises the SM Higgs mass to the observed value without resorting to large SUSY breaking effects in the top/stop sector. This significantly relaxes the pressure on naturalness coming from radiative corrections to the Higgs mass. This leading BMSSM operator further modifies the vacuum and, by supersymmetry, the electroweakino phenomenology in a correlated way. We analysed in this paper various implications of this effective operator, setting its coefficient so as to reproduce $m_h=125\,$GeV at the classical level, {\it i.e.} $-0.1\lesssim\eps_1\lesssim -0.05$ depending on $\tan\beta$ and the CP-odd Higgs scalar mass. 

First of all, we derived at $\mathcal{O}(\eps_1)$ the modifications of the EW fine-tuning associated with tree-level contributions to the weak scale. We further found that {\it for fixed values of the MSSM parameters} the BMSSM correction always suppresses the dominant sources of EW fine-tuning associated with the $\mu$ parameter and the $H_u$ soft mass term, up to $\simeq 40\%$ for $\mu\simeq 100\,$GeV. The fine-tuning improvement however   is significantly reduced as $m_A$ and/or $\tan\beta$ increase above $\simeq 200\,$GeV and $\simeq 4$, respectively.  

Under the assumption that the observed DM is a stable neutralino relic, present direct searches already strongly constrain the LSP composition to either quasi-pure gaugino or Higgsino states, with purity $p={\rm min}(F_{\tilde H},1-F_{\tilde H})\lesssim 0.1$ for DM lighter than $100\,$GeV. This results in a significant source of pressure on naturalness for gaugino DM, since the higher gaugino purity pushes the $\mu$-parameter to increasingly large values. We showed that  the Higgs coupling to gaugino LSP pairs, which controls the SI DM scattering off nucleons, is always enhanced by  the BMSSM contribution at $\mathcal{O}(\eps_1)$. Furthermore the leading MSSM-like contribution to this coupling, which scales like $\tan^{-1}\beta$, cannot be as small as in the MSSM in the presence of the effective correction to the Higgs mass, as the latter strongly favors low $\tan\beta\lesssim 10$. This results in a significantly larger Higgs-to-LSP pairs coupling in the BMSSM for a given value of the $\mu$-parameter, most notably at DM masses below $100\,$GeV. Consequently, the LUX experiment currently implies a minimal fine-tuning as strong as a few percent for a DM mass around $30-50\,$GeV, which is about four times worse than that of the MSSM in the same mass region. Null results from a forthcoming LZ experiment would push the BMSSM fine-tuning to at least the permil level for DM masses above $20\,$GeV, while the MSSM could still be significantly less fine-tuned, up to a factor of $\mathcal{O}(10)$ for DM around $50\,$GeV.  

Direct searches do not constitute, however, an immediate threat to EW naturalness for Higgsino DM, since 
it is sufficient for EW gauginos  to emerge around the scale of a few TeV. Quasi-pure Higgsino DM however suffers from very efficient (co-)annihilation into weak gauge bosons in the early Universe. Whenever kinematically accessible, these processes strongly deplete the Higgsino thermal relic density at freeze-out far below the level required by CMB data, unless the LSP is sufficiently heavy. In the MSSM, this implies $\mu\simeq1\,$TeV and in turn a permil level tuning. This conclusion still holds in the presence of the BMSSM operator as its effect, of $\mathcal{O}(\eps_1v^2/\mu^2)$, is negligible for DM masses around the TeV scale. However, we find that the BMSSM operator marginally allows for a low fine-tuning scenario where the Higgsino DM is just below the weak boson annihilation threshold, without conflicting with LEP constraints on light charginos and present data from direct searches. The right relic abundance, however, comes only at the price of a few specific features, namely a DM mass around $\simeq75\,$GeV, a  sub-TeV scale wino and a large radiative mass splitting among the neutral and charged Higgsino. These requirements signal a sensitivity of the relic density to fundamental parameters which we estimate to be around one part in ten. Albeit its apparent fragility, we still find this scenario worthy of consideration as it is the only island of naturalness in the BMSSM framework which resists present DM constraints. Nonetheless, any mild improvement in searches either for DM at underground detectors or for charginos at colliders~\cite{Han:2013usa,Baer:2014kya,Han:2014kaa} would be sufficient to wipe it out.
With the exception of the aforementioned peculiar region of parameter space, we find that any solution to the little hierarchy problem in SUSY which involves a heavy supersymmetric extension of the MSSM still suffers from a severe fine-tuning problem, in some cases worse than in the MSSM, if this theory is to explain DM-related observations through a stable neutralino. Therefore, DM considerations seem to favor non-minimal realizations of SUSY with light new degrees of freedom or, eventually, scenarios where a significant fraction of DM does not consist of neutralinos. With the currently advertised prospects for improved sensitivities to WIMP DM in future direct searches, this ``little neutralino DM problem'' in SUSY might surpass the one associated with stop searches at the LHC.

\section*{Acknowledgements}

We thank Fawzi Boudjema, Gilad Perez and Alexander Pukhov for helpful discussions. 
This work was supported in part by the French ANR, Project DMAstroLHC, ANR-12-BS05-006 and by
the {\it Investissements d'avenir}, Labex ENIGMASS. AG is supported by the New Frontiers program of the Austrian Academy of Sciences.
\\
\appendix 

\section{Neutral Higgs spectrum}\label{Hspectrum}

We present here the corrections to the spectrum and mixing angle of the neutral CP-even Higgs states in the presence of the effective operators in Eqs.~\eqref{ops1} and~\eqref{ops2}.
The neutral mass-squared matrix of the CP-even neutral Higgs sector is (in the $h_d^0$, $h_u^0$ basis)
\beq
\mathcal{M}_h^2&=&\left(\begin{array}{cc}m_Z^2c^2_\beta+m_A^2s^2_\beta & -(m_Z^2+m_A^2)\frac{s_{2\beta}}{2}\\
-(m_Z^2+m_A^2)\frac{s_{2\beta}}{2}& m_Z^2s_\beta^2+m_A^2c_\beta^2
\end{array}\right)\nonumber\\
&&+4v^2\left(\begin{array}{cc}
\eps_2s^2_\beta-\eps_1s_{2\beta} & -\eps_1 \\ 
-\eps_1 & \eps_2c^2_\beta-\eps_1s_{2\beta}
\end{array}\right)\,,
\eeq
where  $c_\beta\equiv \cos\beta$, $s_\beta\equiv \sin\beta$, etc and the CP-odd scalar mass is related to the Lagrangian parameters through $m_A^2=(2b+4\eps_1v^2)/s_{2\beta}-4\eps_2v^2$. The light ($h$) and heavy ($H$) eigenstates
are obtained through the orthogonal transformation
\beq\label{cpevenstates}
\left(\begin{array}{c} h_u^0 \\ h_d^0 \end{array}\right) = \left(\begin{array}{c} v\sin\beta \\ v\cos\beta \end{array}\right)+\frac{1}{\sqrt{2}}\left(\begin{array}{cc} \cos\alpha & \sin\alpha \\ -\sin\alpha & \cos\alpha \end{array}\right)\left(\begin{array}{c} h \\ H \end{array}\right)\,.
\eeq
To leading $\mathcal{O}(\eps_1)$, the tree-level masses are (provided $m_A>m_Z$) 
\beq\label{Hmass}
m_{h,H}^2&=&\frac{1}{2}\left(m_Z^2+m_A^2\mp\sqrt{\Delta_h}\right)\nonumber\\
&&+2v^2\left[\eps_2\left(1 \pm c_{2\beta}^2\frac{m_Z^2-m_A^2}{\sqrt{\Delta_h}}\right)-2\eps_1s_{2\beta}\left(1\pm \frac{m_A^2+m_Z^2}{\sqrt{\Delta_h}}\right)\right]\,,
\eeq
with $\Delta_h\equiv m_A^4+m_Z^4-2m_A^2m_Z^2c_{4\beta}$, 
while the mixing angle $\alpha$ relates to the tree-level masses as  
\beq
\frac{\tan2\alpha}{\tan2\beta} = \frac{m_A^2+m_Z^2-\delta_t}{m_A^2-m_Z^2}\,,\quad \delta_t = -8\frac{\epsilon_1v^2}{\sin2\beta}\,;
\eeq
and
\beq
\frac{\sin2\alpha}{\sin2\beta} = -\frac{m_H^2+m_h^2-\delta_s}{m_H^2-m_h^2}\,,\quad \delta_s = \delta_t\left(1+\sin^22\beta\right)\,,
\eeq

Since $\sin 2\beta > 0$,  $\delta_{s,t} >0$ for $\eps_1 < 0$ as required by a large tree-level SM-Higgs mass. In the MSSM, $\tan\beta>1$ implies $\sin 2\beta > 0$ and $\cos 2\beta < 0$, which yields (provided $m_A > m_Z$)
\beq
{\rm MSSM: }\quad \sin2\alpha<0\,,\ \cos2\alpha>0\,, 
\eeq
and $\alpha$ is restricted to the lower-right quadrant: $-\pi/2<\alpha <0$. In the BMSSM, however two combinations of signs can arise
\beq\label{alphaBMSSM}
{\rm BMSSM:} \quad \Big\{\begin{array}{cc}\sin2\alpha>0\,,\ \cos2\alpha>0\,, & {\rm for}\  m_A^2+m_Z^2-\delta_t<0\,; \\ 
\sin2\alpha<0\,,\ \cos2\alpha>0\,, & {\rm for}\  m_A^2+m_Z^2-\delta_t>0\end{array}
\eeq
$\alpha>0$ can be achieved in the large $\tan \beta$ limit where $\delta_t \approx 4 | \eps_1 | v^2 \tan\beta$ provided $m_A$ is not too large. Saturating the condition $|\epsilon_1|\tan\beta\lesssim 1$ yields $\alpha>0$ provided $m_A\lesssim 340\,{\rm GeV}$.
In the decoupling limit, $m_A\gtrsim m_Z$, we have that $\beta-\alpha\simeq \pi/2$. 
\section{Neutralino masses and mixings to $\mathcal{O}(m_Z)$} \label{app:Zneu}

We perform in this section the approximate diagonalization of the neutralino matrix up to $\mathcal{O}(m_Z)$ in the presence of the SUSY-preserving effective operator of Eq.~\eqref{ops1}.
The neutralino mass matrix is 
 \beq\label{Mchi0matrix}
\mathcal{M}_{\chi_0} =\left( \begin{array}{cccc}
M_1 & 0 & -m_Zs_Wc_\beta & m_Z s_Ws_\beta \\
0 & M_2 & m_Zc_Wc_\beta & -m_Zc_Ws_\beta  \\
-m_Z s_W c_\beta & m_Zc_Wc_\beta & 2\eps_1\frac{v^2}{\mu}s^2_\beta & -\mu+2\eps_1\frac{v^2}{\mu}s_{2\beta} \\
m_Zs_Ws_\beta & -m_Zc_Ws_\beta & -\mu +2\eps_1\frac{v^2}{\mu}s_{2\beta}& 2\eps_1\frac{v^2}{\mu}c^2_\beta\end{array}\right)\,.
 \eeq
  It proves useful to diagonalize the Higgsino 2$\times$2 block through the nearly maximal rotation of angle $\theta_{\tilde h}=\pi/4+\delta\theta_{\tilde h}$ 
\beq\label{HinoRotate}
\left(\begin{array}{c} \tilde h_d^0 \\ \tilde h_u^0\end{array}\right) = \left(\begin{array}{cc}
\cos\theta_{\tilde h} & \sin\theta_{\tilde h}\\ -\sin\theta_{\tilde h} & \cos\theta_{\tilde h}\end{array}\right)\left(\begin{array}{c} \tilde h_1^0 \\ \tilde h_2^0\end{array}\right)\,,
\eeq 
with $\delta\theta_{\tilde h} \simeq\eps_1c_{2\beta}\, v^2/(2\mu^2)$.

Consider the limit of a decoupled $\tilde W$, as motivated by constraints from direct DM searches ($g>g'$). The neutralino mass matrix of Eq.~\eqref{Mchi0matrix} then reduces to (in the $\tilde B$, $\tilde h_{1}^0$, $\tilde h_{2}^0$ basis)
\beq\label{MnoW}
\mathcal{M}_{\chi_0}^{M_2\to\infty} &\simeq &
\left(\begin{array}{ccc} 
M_1  & - \frac{m_Z s_W(s_\beta+c_\beta)}{\sqrt{2}}\left(1-\delta_-\right)&\frac{m_Zs_W(s_\beta-c_\beta)}{\sqrt{2}}\left( 1+\delta_+\right) \\
\cdot  & \mu_+ & 0 \\
\cdot &\cdot & -\mu_-
\end{array}\right)\nonumber\\
&&-\frac{m_W^2}{2M_2} \left(\begin{array}{ccc} 
0 & 0 & 0\\
\cdot & 1+s_{2\beta}-\delta_0 & c_{2\beta}\left(1+\delta_+-\delta_-\right) \\
\cdot & \cdot  & 1-s_{2\beta} + \delta_0
\end{array}\right)+\mathcal{O}(M_2^{-2})\,.
\eeq
where $\cdot$'s denote entries obtained through the symmetry property of $\mathcal{M}_{\chi_0}$ and
\beq
\mu_\pm\equiv \mu\left(1\pm3\delta_\mp\mp\delta_\pm\right)\,,
\eeq
with
\beq
\delta_\pm \equiv (1\pm\sin2\beta)\frac{ \eps_1v^2}{2\mu^2}\,,\quad  \delta_0\equiv\cos^22\beta\frac{\eps_1v^2}{\mu^2}\,.
\eeq
When $|M_1|$, $\mu$ and their difference are much larger than $m_Z$, the off-diagonal entries in Eq.~\eqref{MnoW} can be treated perturbatively. In this case, the mixing angles between $\tilde B$ and $\tilde h_{1,2}^0$ are approximately
\beq\label{BHangles}
\theta_\pm\simeq \mp\frac{m_Zs_W(s_\beta\pm c_\beta)}{\sqrt{2}(M_1\mp \mu_\pm)} \left( 1\mp\delta_\mp\right)\,,
\eeq 
respectively. We consider below the limiting cases where the LSP is either a nearly pure bino or Higgisno state. 

\subsection{Bino dark matter}

We further assume here $|M_1|\ll\mu$, so that the lightest neutralino $\chi$ is mostly $\tilde B$ with small $\theta_\mp$ projections on $\tilde h_1^0$ and $\tilde  h^0_2$, respectively:
\beq\label{LSPcompoBino}
\chi\simeq \tilde B + \theta_+ \tilde h_1^0 + \theta_-\tilde h_2^0 + \mathcal{O}(\theta_{\pm}^2)\,,
\eeq
where the mixing angles in Eq.~\eqref{BHangles} reduce to
\beq\label{BHangleBino}
\theta_{\pm} \simeq \frac{m_Zs_W(s_\beta\pm c_\beta)}{\sqrt{2}\mu} \left[ 1\pm\frac{M_1}{\mu}+\frac{\eps_1v^2}{2\mu^2}\left(5s_{2\beta}\mp3\right)\right]+\cdots\,,
\eeq
with $\cdots$ denoting neglected $\mathcal{O}(\eps_1M_1/\mu)$ and $\mathcal{O}(M_1^2/\mu^2)$ and higher. Plugging back Eq.~\eqref{BHangleBino} into Eq.~\eqref{LSPcompoBino} and moving back to the original current basis with Eq.~\eqref{HinoRotate} gives the following LSP composition
\beq
\mathcal{N}_{\chi1}\simeq 1\,,\quad \mathcal{N}_{\chi2}\sim \mathcal{O}(M_2^{-1})\,,
\eeq
\beq
\mathcal{N}_{\chi3}&\simeq& \frac{m_Zs_Ws_\beta}{\mu}\left[1+\frac{M_1}{t_\beta\mu}+\frac{\eps_1v^2}{\mu^2}\left(3s_{2\beta}-\frac{2}{t_\beta}\right)\right]\,,\\
\mathcal{N}_{\chi4}&\simeq& -\frac{m_Zs_Wc_\beta}{\mu}\left[1+\frac{t_\beta M_1}{\mu}+\frac{\eps_1v^2}{\mu^2}\left(1+3c_{2\beta}\right)t_\beta\right]\,,
\eeq
which, when used in Eq.~\eqref{hXX} yields the approximate Higgs-to-LSP pair coupling in Eq.~\eqref{hXXbino}.

\subsection{Higgsino dark matter}

We assume here $|M_1|\gg \mu$, so that the lightest neutralino is either of the two Higgsino states $\tilde h_{1,2}^0$ with a small $\theta_\pm$ projection on $\tilde B$. In the MSSM,  $\mathcal{O}(m_Z^2)$ mixings are required to decide which of $\tilde h_{1,2}^0$ is the LSP~\cite{DreesNojiri}, while the degeneracy is dominantly lifted at $\mathcal{O}(\eps_1)$ in the BMSSM. Indeed, the masses of  $\tilde h_1^0$ and $\tilde h_2^0$ are (including $\mathcal{O}(m_Z^2)$ corrections)
\beq\label{ninomasses}
m_{\tilde h_1^0} \simeq \mu_+ +\delta_Z\,,\quad m_{\tilde h_2^0}\simeq \mu_- + \delta_Z'\,,
\eeq
with 
\beq
\delta_Z&\equiv& \frac{m_Z^2}{2}(1+s_{2\beta})\left(\frac{c_W^2}{\mu-M_2}+\frac{s_W^2}{\mu-M_1}\right)\,,\nonumber\\
\delta_Z'&\equiv & \frac{m_Z^2}{2}(1-s_{2\beta})\left(\frac{c_W^2}{\mu+M_2}+\frac{s_W^2}{\mu+M_1}\right)\,,
\eeq
which yields a splitting of
\beq\label{ninosplit}
m_{\tilde h_1^0}-m_{\tilde h_2^0}&\simeq& (\mu_+-\mu_-) +(\delta_Z-\delta_Z')\nonumber\\
&\simeq & 2\frac{\eps_1v^2}{\mu}- \left(\frac{m_Z^2s_W^2}{M_1}+\frac{m_W^2}{M_2}\right)\,,
\eeq 
where  $\mathcal{O}(M_{1,2}^{-2})$ corrections and higher are neglected.
For $\eps_1<0$, we always have $\mu_+<\mu_-$ and $\tilde h_1^0$ is the LSP, unless either of $M_{1,2}$ is negative (for $\mu>0$)  and of sufficiently small magnitude. Assuming {\it e.g.} $M_2\to \infty$, this corresponds to $M_1<0$ and $|M_1|\lesssim m_Z^2s_W^2\mu/(2|\eps_1|v^2)\simeq30\,$GeV for $\mu=100\,$GeV and $|\eps_1|\simeq 0.1$.
 
Focusing for instance on the limit of decoupled wino ($M_2\to \infty$), the LSP is mostly $\tilde h_1^0$ (regardless of the relative sign between $M_1$ and $\mu$)  whenever the BMSSM operator is dominant in raising the Higgs mass above $m_Z$ and the $\mu$-parameter is kept light to minimize fine-tuning. The LSP composition is found to be 
\beq\label{Hinocomp}
\chi \simeq \tilde h_1^0 - \theta_{+}\tilde B+\mathcal{O}(\theta_{\pm}^2)\,,
\eeq 
with 
\beq\label{thetaHino}
\theta_+\simeq -\frac{m_Zs_W(s_\beta+c_\beta)}{\sqrt{2}M_1} \left[ 1-\frac{\eps_1v^2}{2\mu^2}\left(1-s_{2\beta}\right)+\mathcal{O}\left(\frac{\mu}{M_1}\right)\right]\,.
\eeq
or equivalently in the $(\tilde h_d^0,\tilde h_u^0)$ basis
\beq
\mathcal{N}_{\chi1}\simeq -\theta_+\,,\quad \mathcal{N}_{\chi2}\sim \mathcal{O}(M_2^{-1})\,,
\eeq
\beq
\mathcal{N}_{\chi3}\simeq \frac{1}{\sqrt{2}}\left(1-\frac{\eps_1 c_{2\beta}v^2}{2\mu^2}\right)\,,\quad \mathcal{N}_{\chi4}\simeq -\frac{1}{\sqrt{2}}\left(1+\frac{\eps_1 c_{2\beta}v^2}{2\mu^2}\right)\,.
\eeq

An important quantity in Higgsino LSP scenarios is the mass splitting between the LSP and the other neutral and charged Higgsinos. The mass splitting between $\tilde h_1^0$ and $\tilde h_2^0$ is given by Eq.~\eqref{ninosplit}. The charged Higgsino mass is also corrected at $\mathcal{O}(\epsilon_1)$. The chargino mass matrix reads
\beq
\mathcal{M}_{\chi^\pm} =\left(\begin{array}{cc}
M_2 & \sqrt{2}m_W s_\beta \\
\sqrt{2} m_W c_\beta & \mu - s_{2\beta}\frac{\eps_1v^2}{\mu}
\end{array}\right)\,,
\eeq
and the lightest chargino mass in the $M_2\gg \mu$ limit is approximately
\beq\label{cinomass}
m_{\tilde C}\simeq \mu-s_{2\beta}\left(\frac{m_W^2}{M_2}+\frac{\eps_1 v^2}{\mu}\right)\,,
\eeq
modulo neglected $\mathcal{O}(M_2^{-2})$ and higher. Combining Eqs.~\eqref{cinomass},~\eqref{ninomasses} and~\eqref{ninosplit} yields the mass splitting in Eq.~\eqref{charginosplit}.

\section{Spin-independent DM scattering on nucleons}\label{app:SIscattering}
We review here the calculation of the scattering cross section relevant to direct DM searches. The SI cross section on proton (similar expressions can be derived for neutron) is obtained through~\cite{MicrOmegas22}
\beq
\sigma_{\rm SI} = \frac{4m_r^2}{\pi}|f_{p}|^2\,,
\eeq
where the reduced mass $m_r\equiv m_{p}m_\chi/(m_{p}+m_\chi)$ and
\beq\label{fp}
\frac{f_{p}}{m_{p}}= \sum_{q=u,d,s} f_q^{p} A_q + \frac{2}{27}f_g^{p}\sum_{Q=c,b,t} \left(1+\frac{35\alpha_s(m_Q)}{36\pi}\right)A_Q\,,
\eeq
including QCD corrections at NLO~\cite{MicrOmegas22}.
$f_g^{p}= 1-\sum_q f_q^{p}$ and $\alpha_s(m_Q)$ is the running QCD fine structure constant evaluated at the scale $m_Q$. We use~\cite{XingZhangZhou} $\alpha_s(m_c)=0.39$, $\alpha_s(m_b)=0.22$ and $\alpha_s(m_t)=0.108$, and $f_u^p=0.0153$, $f_d^p=0.0191$ and $f_s^p=0.0447$~\cite{MicrOmegas3}.  Assuming universal contributions in the up- and down-type quark sectors, {\it i.e.} $A_u=A_c=A_t\equiv A_{q^u}$ and $A_d=A_s=A_b\equiv A_{q^d}$, Eq.~\eqref{fp} becomes approximately
\beq\label{fpmpapprox}
\frac{f_p}{m_p}\simeq 0.162 A_{q^u} + 0.137A_{q^d}\,.
\eeq

Neglecting squark exchange, the short-distance amplitudes are supported by $t$-channel Higgs exchanges
\beq
A_q \equiv -\frac{1}{2\sqrt{2}v}\left(\frac{g_{h\chi\chi}}{m_h^2}a_q^{h} + \frac{g_{H\chi\chi}}{m_H^2}a_q^H\right)\,, 
\eeq 
with 
\beq
a_{q=u,c,t}^h = \frac{\cos\alpha}{\sin\beta}\,,\quad a_{q=d,s,b}^h = -\frac{\sin\alpha}{\cos\beta}\,,
\eeq
for the light CP-even Higgs boson and
\beq
a_{q=u,c,t}^H = \frac{\sin\alpha}{\sin\beta}\,,\quad a_{q=d,s,b}^H = \frac{\cos\alpha}{\cos\beta}\,,
\eeq
for the heavy one. Recall that in the decoupling limit, $m_A\gg m_Z$, $a_q^h\to 1$ for all $q$, while $a^H_{d,s,b} \to \tan\beta$ and $a^H_{u,c,t} \to -1/\tan\beta$. We checked that cross sections obtained using the above expressions agree with those resulting from {\tt micrOMEGAs} within a percent.\\

Note that in the MSSM with $\tan\beta>1$, since $\pi/2<\alpha<0$ (or, equivalently $\cos\alpha>0$ and $\sin\alpha<0$), the light Higgs contributions to $A_{q^u}$ and $A_{q^d}$ have the same sign and they always add up in the SI scattering cross section. The situation could be rather different in the BMSSM since $\alpha>0$ is possible, see Eq.~\eqref{alphaBMSSM}, in which case $A_{q^u}$ and $A_{q^d}$  interfere destructively. Strong cancellations among the up- and down-type contributions to the light Higgs exchange amplitude can in particular occur for $\tan\beta\sim \mathcal{O}($few) or more and $m_A\lesssim\mathcal{O}(300\,$GeV). In this case, $\beta=\pi/2-\varepsilon_\beta$ and $\alpha=\varepsilon_\alpha>0$, with $\varepsilon_{\alpha}\simeq\varepsilon_\beta\ll 1$, which yields $A_{q^u}\simeq -A_{q^d}$ and in turns leads to potentially strong accidental cancellations in $f_p/m_p$ as given by Eq.~\eqref{fpmpapprox}.
        
\section{Electroweak fine-tuning expressions} \label{app:DeltaEW}

We provide in this section the complete expressions for the individual sources of EW fine-tuning, including $\mathcal{O}(\eps_{1,2})$ corrections, as defined in Eq.~\eqref{FTindiv}. 
\beq
\delta(\mu) = \delta_{\rm MSSM}(\mu)-8\frac{\mu^2}{m_Z^2}
\frac{v^2t_{2\beta}}{m_A^2c_{2\beta}}\left[2\eps_{1}\left(1+s^2_{2\beta}\frac{m_Z^2}{m_A^2}\right)-\eps_{2}s_{2\beta}\left(1 + \frac{2m_Z^2}{m_A^2}\right)\right]\,,
\eeq
\beq
\delta(b) = \delta_{\rm MSSM}(b)+\frac{2\eps_{2}v^2t_{2\beta}^2}{m_Z^2}-\frac{4\eps_{1}v^2s_{2\beta}}{m_A^2}\,,
\eeq
\beq
\delta(m_{H_u}^2) &=& \delta_{\rm MSSM}(m_{H_{u}}^2) -2\eps_1v^2s_{2\beta}\left[\frac{2F_u}{m_A^2}\left[1+\left(1+\frac{m_Z^2}{m_A^2}\right)t^2_{2\beta}\right]-\frac{G_u}{m_Z^2}\right] \nonumber\\
&&+2\eps_2v^2\left[\frac{F_u}{m_A^2}t^2_{2\beta}\left(1+\frac{2m_Z^2}{m_A^2}\right)+\frac{G_u}{m_Z^2}c_\beta^2\right]\,,
\eeq
\beq
\delta(m_{H_d}^2) &=&\delta_{\rm MSSM}(m_{H_{d}}^2) +2\eps_1v^2s_{2\beta}\left[\frac{2F_d}{m_A^2}\left[1+\left(1+\frac{m_Z^2}{m_A^2}\right)t^2_{2\beta}\right]-\frac{G_d}{m_Z^2}\right]\nonumber\\
&&-2\eps_2v^2\left[\frac{F_d}{m_A^2}t^2_{2\beta}\left(1+\frac{2m_Z^2}{m_A^2}\right)+\frac{G_d}{m_Z^2}s_\beta^2\right]\,,
\eeq
\beq
\delta(\eps_{1r})=\frac{8\eps_{1r}v^2}{m_Z^2}s_{2\beta}\left[1+\frac{m_Z^2}{2m_A^2}+\left(1+\frac{m_Z^2}{m_A^2}\right)t^2_{2\beta}\right]\,,
\eeq
\beq
\delta(\eps_{2r})=-\frac{2\eps_{2r}v^2}{m_Z^2}t^2_{2\beta}\left(1+\frac{m_Z^2}{m_A^2}\right)\,,
\eeq
with
\beq
F_u&\equiv& \frac{c_{2\beta}}{2}-\frac{\mu^2}{m_Z^2}+c^2_\beta\frac{m_A^2}{m_Z^2}\,,\\
G_u&\equiv& -1+\frac{1}{c_{2\beta}}-\left(1+\frac{m_Z^2}{m_A^2}\right)t^2_{2\beta}\,,
\eeq
and
\beq
F_d&\equiv&\frac{c_{2\beta}}{2}+\frac{\mu^2}{m_Z^2}-s^2_\beta\frac{m_A^2}{m_Z^2} \,,\\
G_d&\equiv& 1+\frac{1}{c_{2\beta}}+\left(1+\frac{m_Z^2}{m_A^2}\right)t^2_{2\beta}\,,
\eeq
while the MSSM contributions are 
\beq
\delta_{\rm MSSM}(\mu) &=& -\frac{4\mu^2}{m_Z^2}\left(1+\left(1+\frac{m_Z^2}{m_A^2}\right)t^2_{2\beta}\right)\,,\\
\delta_{\rm MSSM}(b) &=& t^2_{2\beta}\left(1+\frac{m_A^2}{m_Z^2}\right)\,,\\
\delta_{\rm MSSM}(m_{H_{u,d}}^2)&=& F_{u,d}\, G_{u,d}\,. 
\eeq

\bibliographystyle{JHEP}
\providecommand{\href}[2]{#2}\begingroup\raggedright\endgroup

\end{document}